\documentclass{article}

\usepackage[main, final]{iaseai26}

\usepackage[utf8]{inputenc} %
\usepackage[T1]{fontenc}    %
\usepackage{hyperref}       %
\usepackage{url}            %
\usepackage{booktabs}       %
\usepackage{amsfonts}       %
\usepackage{nicefrac}       %
\usepackage{microtype}      %
\usepackage{xcolor}         %
\usepackage{float}

\title{Auditing a Dutch Public Sector Risk Profiling Algorithm Using an Unsupervised Bias Detection Tool}

\author{%
  Floris Holstege \\
  University of Amsterdam\\
  Netherlands\\
  \And
  Mackenzie Jorgensen \\
  Northumbria University \\
  United Kingdom \\
   \And
  Kirtan Padh \\
  TU Munich \\
  Germany \\
   \And
  Jurriaan Parie\thanks{Corresponding author: j.parie@algorithmaudit.eu} \\
  Algorithm Audit \\
  Netherlands \\
   \And
  Krsto Prorokovi\'{c} \\
  Algorithm Audit \\
  Netherlands \\
     \And
  Joel Persson \\
  Algorithm Audit \\
  Netherlands \\
     \And
  Lukas Snoek \\
  Algorithm Audit \\
  Netherlands
}

\usepackage[ruled, linesnumbered, noalgohanging]{algorithm2e}
\usepackage{listings}
\usepackage{subcaption}
\usepackage{subcaption}
\usepackage{graphicx}
\usepackage{amsmath}

\newcommand{\protectedgroup}{demographic group}
\newcommand{\barM}{\bar{M}}

\newcommand{\boldxsub}[1]{\boldsymbol{x}_{#1}}

\newcommand{\ysub}[1]{y_{#1}}

\newcommand{\boldmu}{\boldsymbol{\mu}}
\newcommand{\boldSigma}{\boldsymbol{\Sigma}}

\newcommand{\littlespace}{\hspace{0.25cm}}
\newcommand{\nperm}{n_{\mathrm{perm}}}

\begin{document}

\maketitle

\begin{abstract}
Algorithms are increasingly used to automate or aid human decisions, yet recent research shows that these algorithms may exhibit bias across legally protected demographic groups. However, data on these groups may be unavailable to organizations or external auditors due to privacy legislation. This paper studies bias detection using an unsupervised bias detection tool when data on demographic groups are unavailable. We collaborated with the Dutch Executive Agency for Education to audit an algorithm that was used to assign risk scores to college students at the national level in the Netherlands between 2012-2023. Our audit covers more than 250,000 students across the country. The unsupervised bias detection tool highlights known disparities between students with a non-European migration background and students with a Dutch or European-migration background. Our contributions are two-fold: (1) we assess bias in a real-world, large-scale, and high-stakes decision-making process by a governmental organization; (2) we provide the unsupervised bias detection tool in an open-source library for others to use to complete bias audits. Our work serves as a starting point for a deliberative assessment by human experts to evaluate potential discrimination in algorithmic decision-making.
\end{abstract}

\section{Introduction}
\label{sec:intro}

Between 2012 and 2023, a simple rule-based risk profiling algorithm from the Dutch Executive Agency for Education (DUO) contributed to indirect discrimination in a control process to check whether students were unduly allocated a college grant~\citep{DutchParliament2024D07565}. A rule-based algorithm, consisting of three apparently neutral characteristics of the students (type of education, age, and distance to parents), was used to assign higher risk scores to students enrolled in vocational education, younger students, and students who lived near their parental address. The seemingly impartial risk assessment led to unequal treatment of students with a non-European migration background, remaining unnoticed for over 10 years and affecting more than 350,000 students~\citep{DutchParliament24724-240}.~\footnote{Between 2012-2023, more than 350,000 students were assigned a risk score by the algorithm. Our analysis focuses on the student population in two years: 2014 and 2019, which consists of a total of 298,882 students.} ``Risk'' in this context, and throughout the text, refers to the risk of ``undue'' or improper granting of college grants to students.

Given this, in 2024, the Dutch Minister for Education, Culture and Science apologized on behalf of the Dutch government for indirect discrimination experienced by students with a non-European migration background and announced a EUR 61 million compensation plan~\citep{DutchParliament24724-240}. This recent discriminatory algorithm scandal in the Netherlands~\citep{politico22} highlights that not only complex statistical methods, such as machine learning, but also simple rule-based algorithms can pose significant risks of embedding discrimination in algorithmic decision-making processes. In a recent interview, the chair of the Dutch Data Protection Authority warned that discriminatory rule-based algorithms are ``around every corner'' in executive branches of the Dutch government~\citep{VK24}. 
This type of risk profiling is prohibited by the European Convention of Human Rights (ECHR), the Charter of Fundamental Rights of the European Union (EU)~\citep{charter}, the EU General Data Protection Regulation (GDPR)~\citep{gdpr_2016} and the EU AI Act~\citep{AIAct}, as well as the national Public Administration Law~\citep{awb}. Limited institutional enforcement capabilities, along with a lack of awareness, have led to inadequate oversight of algorithmic decision-making processes~\citep{meuwese2024algoprudence}. 

To combat bias in algorithmic systems, the academic community has presented a multitude of ways to define, detect, and mitigate different notions of bias at different points along the algorithmic pipeline~\citep{Veale2018, Barocas2019, Chouldechova2020, Mehrabi2021}. Most bias mitigation methods require data on demographic groups to prevent discriminatory bias. These demographic groups are typically protected attributes under discrimination law, such as race, gender and age~\citep{ashurst2023fairness}.\footnote{The exact protected grounds are enshrined in European non-discrimination law, such as the Treaty on the Functioning of the European Union (TFEU), the EU Charter of Fundamental Rights, and specific directives like the Racial Equality Directive (2000/43/EC) and the Employment Equality Directive (2000/78/EC). Note that some provisions do not mention any particular grounds of discrimination, such as Article 20 of the EU Charter of Fundamental Rights which includes openly formulated clauses such as ``everyone is equal before the law.'' The main protected attributes include: race and ethnicity, gender and sex, religion or belief, disability, age, sexual orientation, nationality, and marital or family status. We will refer to these attributes as \emph{{\protectedgroup}s.}} 
However, data on demographic groups are often not available to organizations or external algorithm auditors~\citep{McKane2021,cdei2023}. %
Unsupervised learning can aid in this challenge by detecting groups in the data for which a given notion of equal treatment is violated, avoiding the need for ex-ante specification of demographic or legally protected groups~\citep{MISZTALRADECKA2021102519}. 

In this paper, we use an unsupervised bias detection tool to re-audit the risk profiling algorithm from DUO for bias. The original audit, conducted by Algorithm Audit in 2023, relied on aggregated demographic data on country of birth and country of origin provided by the Dutch national statistics office, \citet{CBS}, and is referred to as ``supervised bias detection.'' For both audits, DUO provided data processed by their algorithm that is not publicly accessible. By combining results from the supervised and unsupervised audit, we assess whether the clusters identified by the unsupervised bias detection tool, which represent potentially unfairly treated groups, correspond to disadvantaged demographic groups resulting from the supervised analysis. This allows for a comparison between an unsupervised learning approach and a supervised approach. We emphasize that the clustering procedure itself does not rely on the available protected demographic data: we only use the aggregated demographic data to illustrate, post hoc, that the flagged clusters align with negatively affected groups. This additional analysis is therefore a validation of the unsupervised clustering methodology, not a replacement for supervised bias detection. 

In our context, the bias metric\footnote{Throughout our paper, we use the term bias metric to refer to what is commonly called a fairness metric in the algorithmic fairness literature. The bias metric we define reflects the fairness metric of demographic parity~\citep{dwork12}.} requires equal probabilities of ``high risk'' classification for students with a Dutch or European migration background and those with a non-European migration background. 
We justify the use of this metric since it assesses the adverse effect of the risk profiling algorithm on different demographic groups: a crucial aspect for investigating indirect discrimination under European non-discrimination law.
We find that the cluster with the highest bias returned by the unsupervised bias detection tool is characterized by students who are more often enrolled in vocational education, live close to their parental address, and are relatively more often 15-18 or 25-50 years old. This aligns with demographic groups found to be negatively affected by the risk profiling algorithm, based on aggregated statistics regarding students' non-European migration background. 

Our paper makes two contributions: (1) A \textbf{large scale audit of indirect discrimination}: The DUO audit presents a high-quality and socially relevant dataset, providing insights into the assessment of indirect discrimination in algorithmic decision-making. This study thus contributes to the advancement of public knowledge on auditing algorithmic systems for indirect discrimination. 
(2) An \textbf{open-source tool}: The unsupervised bias detection method is available as an open-source Python package\footnote{\url{https://github.com/NGO-Algorithm-Audit/unsupervised-bias-detection}} and a web application.\footnote{\url{https://algorithmaudit.eu/technical-tools/bdt/\#web-app}} The latter allows non-technical users to apply the tool to their own datasets in order to detect potential bias. The underlying algorithm we provide includes methodological improvements that substantially reduce the false positive rate.

Overall, our paper showcases the potential of unsupervised learning to aid in auditing algorithmic decision-making processes through a real-world use case. We argue that unsupervised bias detection serves as a starting point for a deliberative assessment by human experts to evaluate potential discrimination and unfairness in the process when demographic groups are unavailable.

\section{Literature review}
\label{sec:lit}

We review a selection of the literature on bias detection and non-discrimination law, focusing on three topics: (1) challenges in detecting and establishing indirect discrimination, (2) bias detection methods with and without access to demographic data, and (3) auditing algorithmic systems for bias.

\subsection{Challenges in detecting and establishing discrimination}

Under European non-discrimination law, DUO's control process is classified as indirect discrimination because a neutral practice disproportionately disadvantaged non-European migrant students~\citep{BilkaKaufhaus}. In algorithms, discrimination often arises from the \emph{proxy and correlation challenge}~\citep{EC_algorithmicdiscrimination}. Proxies refer to characteristics that are correlated with a {\protectedgroup}, leading to disadvantaged outcomes for this group. For example, the Court of Justice of the European Union (CJEU) found that using part-time employment as a basis for differentiation in the context of occupational pension schemes constituted indirect discrimination, since more than 80\% of part-time workers in Spain in 2012 were women~\citep{MorenoInstitutoNacional}. 
In some cases, courts may accept that differential treatment of a {\protectedgroup} is acceptable. Under EU law, indirect discrimination can be justified if it is demonstrated that a legitimate aim is being pursued, and the means of achieving that aim are appropriate and necessary~\citep{RacialEqualityDirective,EmploymentEqualityDirective,GoodsandServicesDirective,RecastGenderEqualityDirective}. Our paper highlights how unsupervised methods can aid in the challenge of assessing indirect discrimination in algorithmic decision-making processes when demographic group data is not available.

\subsection{Bias detection with and without access to demographic groups}
In the algorithmic fairness literature, identifying systematic differences in the treatment of individuals or groups compared to others is commonly referred to as \emph{bias detection}. Most bias detection methods assume access to demographic groups~\citep{Elliot2009, Hardt2016, Kusner2017, Kleinberg2018,Aif360,delaney2024oxonfair}. However, in practice, these labels are often unavailable due to privacy legislation~\citep{van2023using}. For instance, Article 9 in the EU's GDPR restricts the collection of ``special categories of data''~\citep{gdpr_2016}, which include data on ethnicity, religion, health, and sexual orientation.\footnote{Age and gender are not considered special categories of data, see Article 9(1) GDPR. Although, a caveat could exist~\citep{van2018trans}. The other way around, ``political opinions,'' ``trade union membership,'' ``genetic,'' and ``biometric'' data are special categories of data but are not protected by EU non-discrimination directives.} The EU AI Act introduces a potential exception to the prohibition on processing special category data: ``to the extent that it is strictly necessary for the purpose of ensuring bias detection and correction''~\citep{AIAct}. Nonetheless, ongoing legal uncertainties regarding the interplay between the AI Act and the GDPR underscore the need for methods to assess bias in algorithmic systems without relying on demographic data. Algorithms can be biased even in the absence of demographic data, particularly through proxy discrimination, where a set of attributes acts as a proxy for the protected attribute~\citep{ashurst2023fairness}. %

Few alternatives exist to approaches that rely on demographic group data. One class of methods circumvents the limited availability of such data by adopting a Rawlsian definition of fairness~\citep{rawls2001justice} that maximizes utility for the most disadvantaged group~\citep{ hashimoto2018fairness, lahoti2020fairness, chai2022self}. However, the results of these methods do not always adhere to more traditional parity-based group fairness metrics~\citep{islam2024fairness}. A set of alternatives are unsupervised learning methods for fairness~\citep{Nasiriani_2019, 2021BMVC_UDIS}. These methods tend to modify traditional clustering algorithms to handle fairness criteria explicitly. Our work contributes to this latter category. %
Unsupervised learning methods are particularly helpful to examine the effects of apparently neutral provisions that might result in indirect discrimination when demographic groups are not known beforehand.

\subsection{Auditing algorithmic systems}
Inspired by established practices in non-algorithmic disciplines, auditing has the potential to mitigate ethical and legal risks in algorithmic systems through both internal and external oversight~\citep{raji2022outsider}. Legislators across various jurisdictions are implementing regulations to impose audit frameworks for these systems. For instance, since 2023, New York City has required automated employment decision-making tools to undergo audits by independent third-party auditors~\citep{LL144}. Similarly, since 2024, the EU Digital Services Act mandates annual independent third-party audits for online platforms and search engines with more than 45M annual users~\citep{DSA}. In the coming years, the EU AI Act places predominantly self-regulatory obligations on producers and deployers of high-risk AI systems~\citep{AIAct}. 

Implementing effective audits for algorithmic systems poses practical and cultural challenges. Auditors frequently encounter obstacles such as restricted data access, differing views on what constitutes a legitimate auditor~\citep{groves24}, and uncertainty about which fairness metrics best fit the context~\citep{corbett2023measure}. Furthermore, auditing reports are not always required to be disclosed, reducing transparency and hindering the development of public knowledge regarding auditing standards. By discussing the audit of DUO's risk profiling algorithm, we contribute to public knowledge on best practices for auditing algorithmic systems.

\section{Insights into the unsupervised bias detection tool}\label{sec:tool}
\begin{figure*}[t]
     \centering
     \includegraphics[width=0.75\linewidth]{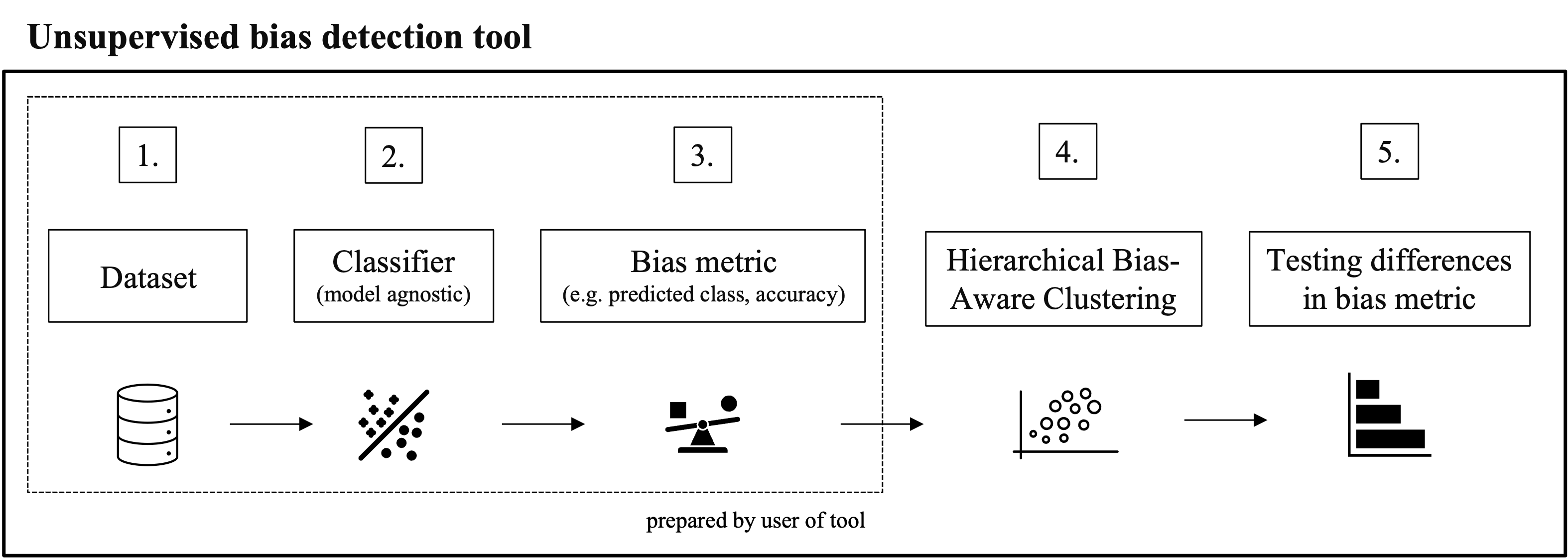}
\caption{Schematic overview of the steps involved in applying the unsupervised bias detection tool. The required information is a dataset, a classifier's outputs, and a bias metric. 
    Part of the dataset is used to train the Hierarchical Bias-Aware Clustering algorithm \citep{MISZTALRADECKA2021102519}. Another part of the dataset is used to test whether differences in the bias metric across clusters are statistically significant.}
     \label{fig:overall}
\end{figure*}

This section introduces the theoretical background of the unsupervised bias detection tool. We use examples from our DUO case study to clarify the methodology. We highlight that our research provides a realistic case-study of applying the tool to investigate its usefulness when one lacks protected attribute data. Our study is of both academic and practical interest since the initial algorithm is easy to apply but lacked justification, and because our data is both comprehensive and from a high-stakes policy use-case. Figure \ref{fig:overall} provides an overview of all the steps in applying the tool. We also provide a brief overview of our second contribution, the open source bias detection tool. 

\subsection{Notation and problem description}
\label{sec:subsec-notation}
Suppose that we have a total of $N$ students in the dataset, labeled $u_1, u_2, \ldots u_N$. Each row in the dataset reflects characteristics $\boldxsub{i} \in \mathbb{R}^d$ (e.g., education level, age, distance to parents) of a student $u_i$ receiving a grant. Suppose also that we have a risk profiling algorithm $f: \mathbb{R}^d \rightarrow \{0, 1\}$ (e.g., the risk profiling algorithm used by DUO) that takes the features of a user as input and outputs whether a user should be deemed high risk of unduly being given a grant. There exists a true value of this variable of interest, denoted $y_i$, denoting whether the grant was, in fact, duly given (or not). 

We measure the degree to which an algorithm is biased by using a bias metric $M$. For a user $u_i$, the value of the metric $m_i = M(f(\boldxsub{i}), \ysub{i})$ is some function of the prediction $f(\boldxsub{i})$ and (potentially) $\ysub{i}$. To give some concrete examples, this could be a measure of performance, such as accuracy (requiring $f(\boldxsub{i}), \ysub{i}$). Alternatively, it could measure demographic parity (requiring only $f(\boldxsub{i})$), equality of opportunity, etc. Without loss of generality,  we assume that a higher average value for $m_i$ for one group compared to another group indicates bias towards the former group.

For $k \subset \{1, 2,\ldots,N\}$, let $U_k = \{u_i\}_{i \in k}$ be a subset of all applicants $U$. We denote by $\barM(U_k)$ the mean of the metric $m_i$ among all in $U_k$. Our goal is to test whether the difference between $\barM(U\backslash U_k)$ and $\barM(U_k)$ is statistically significant. If we had access to a demographic group for which we would like to investigate bias, we would simply define the subset $U_k$ to be, for example, different genders, races, or intersections thereof. However, since the demographic group is unavailable in our setting, we leverage clustering algorithms to separate the data into subsets.  Let $\mathcal{K} = \{1, ..., K\}$ denote the set of clusters, and let $\boldxsub{i, k}, y_{i, k}, m_{i, k}$ denote the data associated with user $u_i$ belonging to cluster $k \in \mathcal{K}$.

\subsection{The clustering algorithm}
In this paper, we use the Hierarchical Bias-Aware Clustering methodology from \citet{MISZTALRADECKA2021102519}, hereafter referred to as ``HBAC.'' HBAC is an iterative clustering algorithm that produces a partition of the original dataset to isolate potentially unfairly treated groups into distinct clusters. We use HBAC as a starting point for the DUO audit, as it represents a standard method for unsupervised bias detection in recommendation systems \citep{deldjoo2024fairness}. Additionally, we propose methodological enhancements to refine the algorithm for real-world algorithmic auditing.

The HBAC algorithm first considers all observations in the dataset to belong to a single cluster and then iteratively splits the observations into distinct clusters. At each iteration, the algorithm identifies the cluster with the highest standard deviation of a specified bias metric, which serves as an indicator of imbalance between clusters. This cluster is then split into two subclusters. The process repeats for a predefined number of iterations, $max\_iterations$. The pseudocode for the algorithm is provided in Algorithm \ref{alg:hbac} in Appendix \ref{sec:appendix-hbac-algo}. Compared to the original algorithm from \citet{MISZTALRADECKA2021102519}, the proposed method supports $k$-modes (enabling clustering over categorical variables) alongside $k$-means.

The original algorithm from \citet{MISZTALRADECKA2021102519} does not specify how to assign clusters to data points that are not part of the original training dataset. We propose using the centroids of each cluster. The datapoint is assigned to the cluster whose distance from the centroid is the smallest. In the case of $k$-means and $k$-modes, the centroids are defined as, respectively, the mean and mode for each of the characteristics $\boldxsub{i, k}$ \citep{Huang_1998, James2023}. The Euclidean distance ($k$-means) or Hamming distance ($k$-modes) is used to determine which centroid is closer.

\subsection{How to test differences in bias between clusters}
\label{sec:sub-section:stat-testing}
In the original paper, no detailed methodology is outlined for testing whether there is a statistically significant difference in the bias metric between the clusters identified by the HBAC algorithm. It is not immediately evident that the most deviating cluster, with respect to the bias metric, is statistically significantly different from data points outside this cluster. To address this, we formulate a null hypothesis that assumes no difference in bias between the most deviating cluster and the rest of the dataset. Using a bootstrapped null-distribution, we test for differences between clusters, allowing for the construction of a valid statistical hypothesis test. We further suggest, for appropriate statistical testing, to apply sample splitting and adjustments for multiple hypothesis testing. We define the aforementioned tests in Appendix \ref{sec:tests}. Following that, in Appendix \ref{sec:addendum_sim}, we illustrate the test results showing that the combination of the Bonferroni correction~\citep{hochberg1987multiple} and sample splitting prevents us from wrongly concluding that there is a difference in the bias metric while there is none. Thus, we propose methodological improvements to the original HBAC algorithm that minimize false bias detection when using cluster analysis.

\subsection{The open-source bias detection tool}
The unsupervised bias detection tool is available on Github.~\footnote{\url{https://github.com/NGO-Algorithm-Audit/unsupervised-bias-detection}} It implements the HBAC algorithm for detecting potentially discriminated user groups, along with the methodological improvements described in Section \ref{sec:sub-section:stat-testing}.
Currently, the library supports \textit{k}-means and \textit{k}-modes algorithms for splitting the data into clusters. The modular design of the tool ensures that additional splitting methods can be integrated with minimal effort. To enhance accessibility, the tool is embedded within a web application,~\footnote{\url{https://algorithmaudit.eu/technical-tools/bdt/\#web-app}} enabling people with no programming background to use it.

\section{Experiment setup: applying the bias detection tool to DUO's risk profiling}
\label{sec:DUO}

We were granted access to ground truth labels for demographic groups in the form of aggregated statistics on students' migration backgrounds from~\citet{CBS}. This case study is therefore well-suited to evaluate the effectiveness of the unsupervised bias detection tool in detecting bias from DUO's risk profiling algorithm. If these ground-truth group-level data had not been available, the unsupervised bias detection tool discussed in our paper could have served as a method to examine \emph{potential} bias in the DUO  verification of whether a student lived at their stated address. %

\subsection{Description of the DUO case study}\label{sec:dataset}
From 2012 to 2023, DUO aimed to verify that college grants were allocated to students who genuinely lived away from their parents. Some students claimed to have an independent living status to qualify for a grant, even though they did not actually reside at the address they provided~\citep{DutchParliament2024D07565}. The process used to verify whether a student lived at the stated address is known as the \emph{CUB process} (Controle Uitwonendenbeurs) which translates to ``Check Non-Resident Grant.'' As part of the CUB process, students were assigned a risk score ranging between 0 and 180 by a rule-based algorithm (further details are in Appendix~\ref{sec:Appendix_Risk_profile}). This score was subsequently binned into six risk categories: very high (1), high (2), average (3), low (4), very low (5) to unknown (6).\footnote{Risk category 6 includes students with a risk score of 0. Risk categories 5, 4, 3, 2, and 1 correspond to students with risk scores in the ranges of 1-19, 20-39, 40-59, 60-79, and 80-180 respectively (see Appendix~\ref{sec:Appendix_Risk_profile}).} Students in categories (1) and (2) were considered ``high risk'' of being unduly allocated the grant. Based on the assigned risk category, human analysts of DUO followed operational instructions to subject students to an investigation. %
Although students from all risk categories could potentially be investigated, the majority of investigations focused on higher-risk categories~\citep{DutchParliament2024D07565}. 

After concerns were raised regarding potential bias in the CUB process affecting students with a non-European migration background, DUO requested Statistics Netherlands to provide aggregated statistics~\footnote{Through protocols established by Statistics Netherlands, the aggregated population statistics can be computed securely. Data were analyzed in a privacy-preserving manner within a secure environment. Groups are rounded to the nearest ten. Results pertaining to groups smaller than 10 people are not published.} on students' countries of origin and countries of birth \footnote{Students with a non-European migration background are defined as those who were either born outside the Netherlands and Europe~\citep{CBScountries}, or who have at least one parent born outside these regions.} who received a college grant for the years 2014, 2017, 2019, 2021, and 2022~\citep{CBS}. From these data, the non-European migration background of students can be retrieved in aggregate. We do not have individual-level data on students' migration backgrounds.

This paper focuses on the years 2014 and 2019, covering a total of 298,882 students who received college grants. We specifically focus on the CUB population from 2014 ($n=248,649$) as this was the last year all types of students---MBO 1-2 and MBO 3-4 (vocational education), HBO (higher professional education), and WO (university education)---were awarded a college grant. We do not combine data from the different years since in 2019 only MBO students were eligible for a college grant. The results are replicated for the CUB 2019 dataset ($n=50,233$ students) in Appendix \ref{sec:CUB-2019}. 

DUO's risk profiling algorithm differentiated on the basis of three variables. First, the type of education the student is following, categorized as MBO 1-2, MBO 3-4 (vocational education), HBO (higher professional education), and WO (university education). Second, the age of the student, categorized as: 15-18, 19-20, 21-22, 23-24, and 25-50 years. Third, the distance between the address of the parents and the student, measured through the 6-digit postal code of the registered addresses and categorized as: 0km, 1m-1km, 1-2km, 2-5km, 5-10km, 10-20km, 20-50km, 50-500km, and unknown. We note that 0km indicates that the student and their parent(s) share the same postal code but not the same house number.

In 2014, 34,050 students were assigned the risk category 6 (unknown), e.g., because the distance to their parents was unknown. We exclude these students from the clustering analysis, since the algorithm did not make a prediction on their risk of undue allocation of the grant, and these cases were deprioritized in the control process~\citep{DutchParliament2024D07565}. Clustering is applied to the remaining dataset of $n=214,599$ students. 

\subsection{Applying the unsupervised bias detection tool to the CUB 2014 dataset}

\textbf{Data preprocessing:} We transform the 3 categorical profiling characteristics to one-hot encoded variables. We define no ordinal relationship between categories. We use the $k$-modes clustering within Algorithm \ref{alg:hbac} as defined in Appendix \ref{sec:appendix-hbac-algo}.

\textbf{Bias metric}: 
As stated before, our metric requires that the probability of being classified as ``high risk'' by the risk profiling algorithm (based on $f(x_i)$) is equal for students with a European migration background and students with a non-European migration background. Thus, if these probabilities are equal, this condition satisfies our bias metric and corresponds to demographic parity~\citep{dwork12}. We consider this to be an appropriate bias metric, as classifications by the risk profiling algorithm contributed to discrimination in the CUB process~\citep{DutchParliament24724-240}. In addition, the assessment of the adverse effect of the risk profiling algorithm on different demographic groups is a crucial aspect to investigate indirect discrimination under European non-discrimination law. An alternative definition of bias could involve examining whether college grants were duly awarded ($y_i$). However, ground truth labels indicating actual outcomes are only available for a small subset of students who underwent manual review by DUO (n=3,179). Consequently, we use predicted risk scores ($f(x_i)$) instead, enabling the application to a substantially larger population (n=298,882). We emphasize that the HBAC algorithm can be used for any bias metric of interest.

\textbf{Hyperparameters}: The minimum number of samples per cluster is selected by iterating over 5,000, 10,000, 20,000, and 30,000 samples and selecting the one which maximizes the average Calinksi-Harabasz index~\citep{calinski1974dendrite},\footnote{This index measures the ratio of the \textit{between} cluster variance over the \textit{within} cluster variance. Intuitively, we want the clusters to have little variation in the bias metric \textit{within} each cluster, but large variation \textit{between} the clusters. This means that the data points are well separated into distinct clusters.} measured via 5-fold cross-validation. We set the maximum number of iterations sufficiently high ($max\_iterations = 1,000$), such that Algorithm \ref{alg:hbac} in Appendix \ref{sec:appendix-hbac-algo} continues until the minimum number of samples per cluster is reached. The initial centroids are determined via the method proposed by \citet{Cao_2009}.

\textbf{Testing procedure:} We split our dataset and reserve 20\% of it for testing the differences in the bias metric per cluster. We fit the unsupervised bias detection tool (and determine hyperparameters) on the remaining 80\% of the data. Since our bias metric is binary, we use a Pearson $\chi^2$-test~\citep{pearson1900x}. 

\section{Bias audit results and discussion of DUO's risk profiling algorithm}
\label{sec:2014-results}

In this section, we outline the findings of our bias audit for the CUB 2014 dataset. We include results from the CUB 2019 dataset in Appendix \ref{sec:CUB-2019}. Through analyzing both CUB datasets, we show that the unsupervised bias detection tool detected bias in the risk profiling algorithm that aligned with the indirect discrimination of non-European migrants found in the initial audit.

\begin{figure*}[t]
 \centering
\begin{subfigure}{.25\textwidth}
  \centering
\includegraphics[scale=0.195]{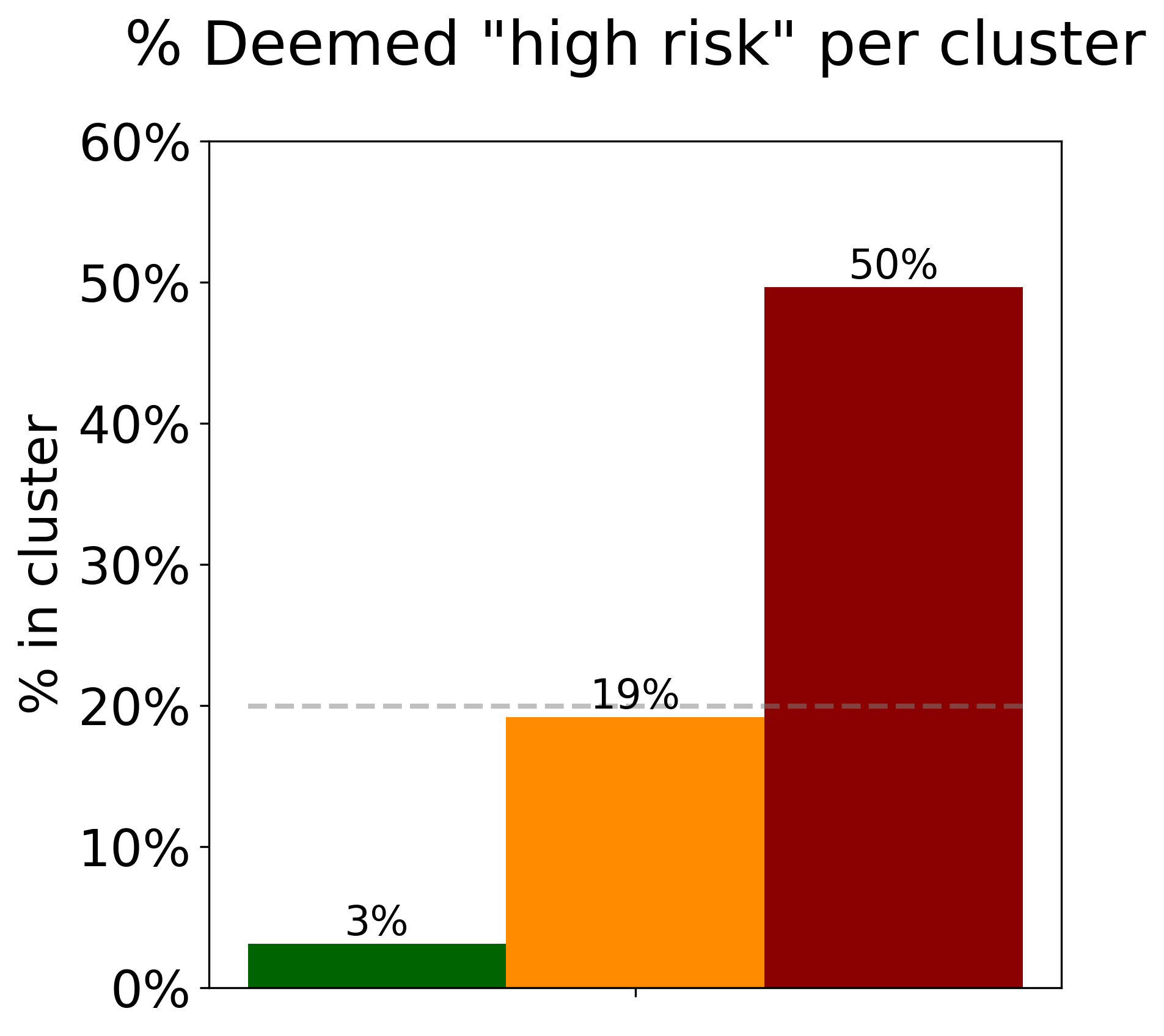}  
\caption{}
\label{fig:predicted_class_cluster}
\end{subfigure}
\begin{subfigure}{.5\textwidth}
  \centering
\includegraphics[scale=0.195]{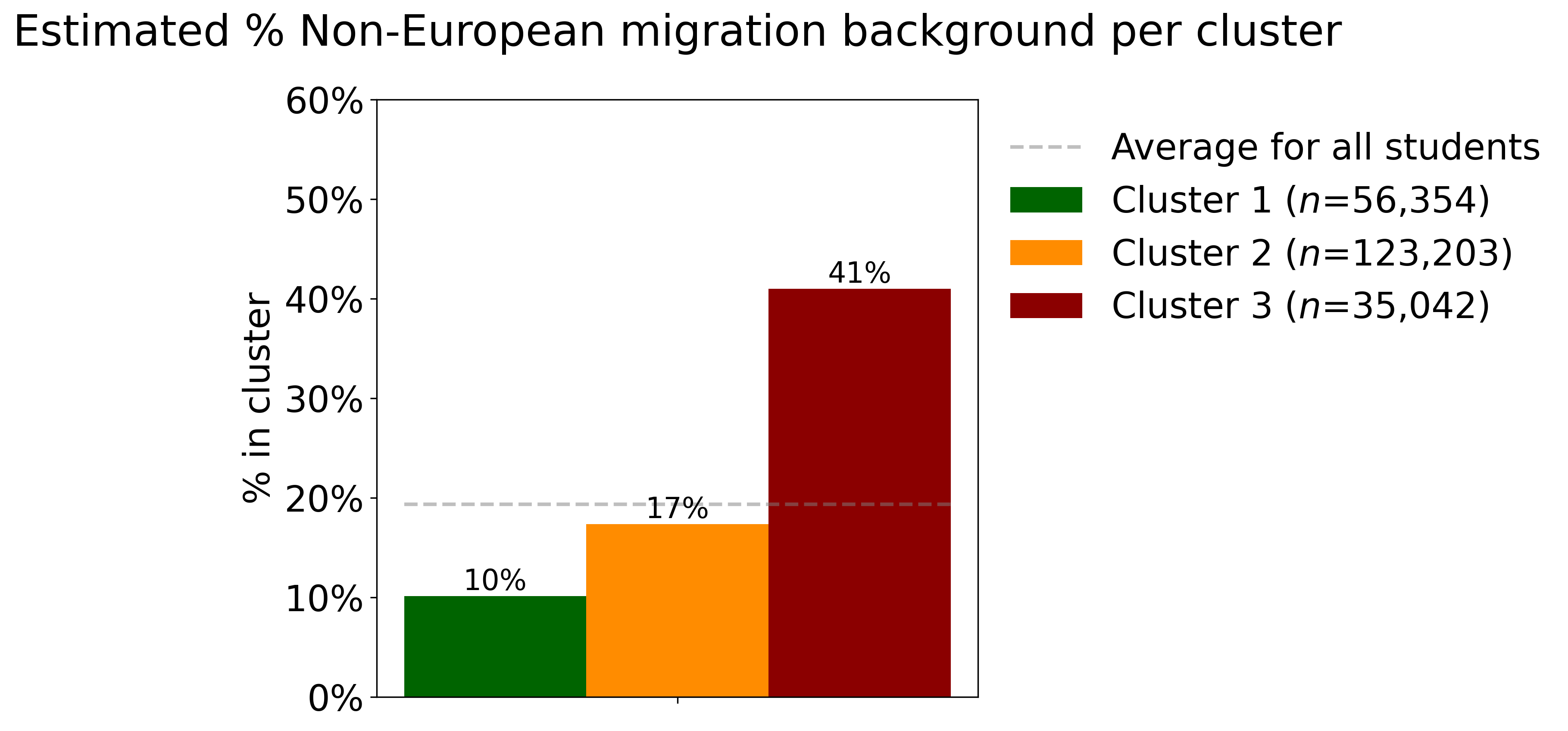}  
  \caption{}
    \label{fig:non_eu_mig_back_cluster}
\end{subfigure}
\caption{The percentage of students (a) deemed as ``high risk'' and (b) estimated as having a non-European migration background for identified clusters based on the student population of 2014, excluding students for which the risk profile was unknown ($n=214,599$). Cluster 1 is green, cluster 2 is yellow, and cluster 3 is red.}
\label{fig:bivariate_bias_background}
\end{figure*}

\begin{figure*}[t]
  \centering
  \begin{subfigure}{.3\textwidth}
    \centering
    \includegraphics[width=\textwidth]{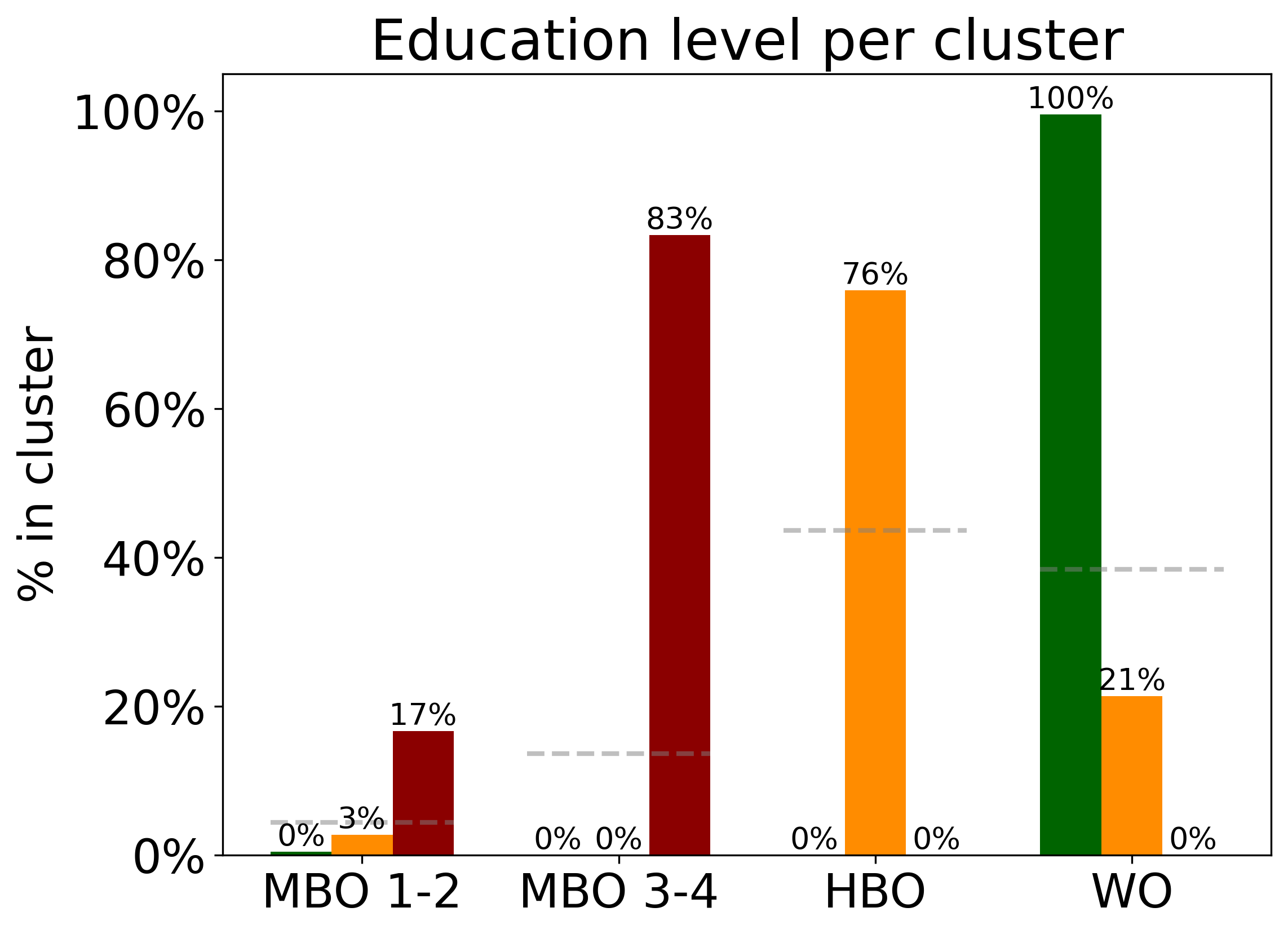}
    \caption{}
  \end{subfigure}
  \hfill
  \begin{subfigure}{.3\textwidth}
    \centering
    \includegraphics[width=\textwidth]{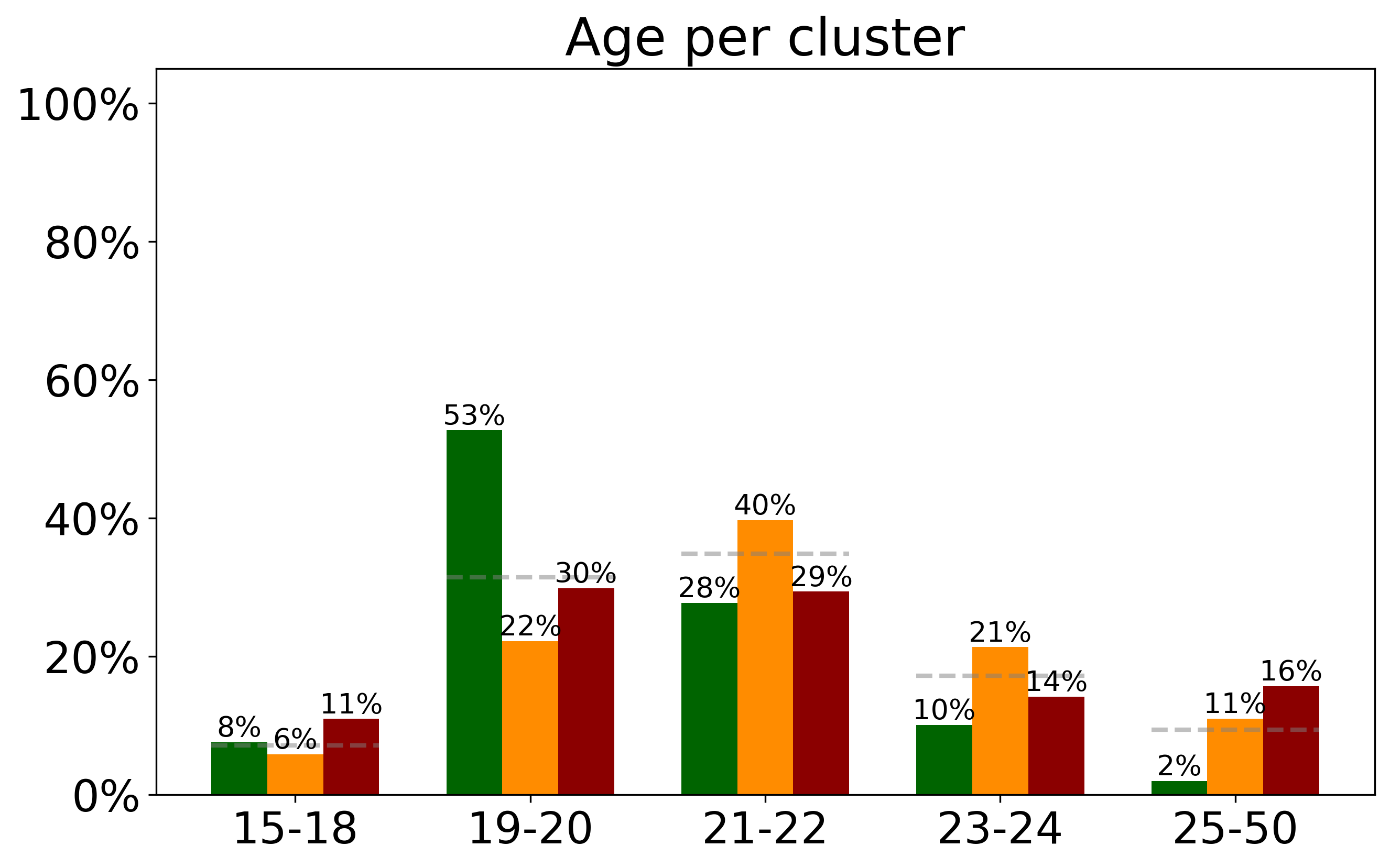}
    \caption{}
    \label{fig:cluster_age}
  \end{subfigure}
  \hfill
  \begin{subfigure}{.3\textwidth}
    \centering
    \includegraphics[width=\textwidth]{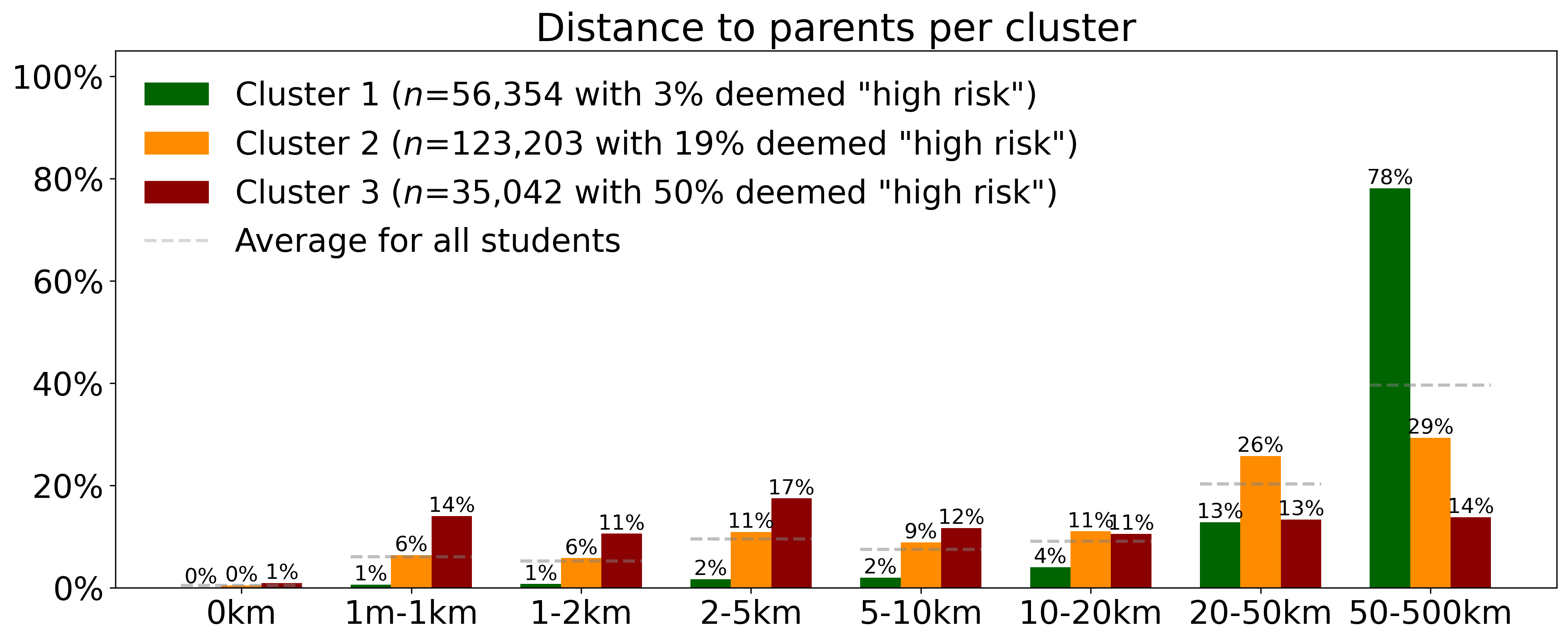}
    \caption{}
     \end{subfigure}
  \caption{The percentage of students in each cluster for the three characteristics used in the risk profiling algorithm: (a) type of education, (b) age, and (c) distance to parents within the student population of 2014, excluding students for which the risk profile was unknown ($n=214,599$). For each subgroup, the average is indicated by a dashed line. Cluster 1 is green, cluster 2 is yellow, and cluster 3 is red.}
  \label{fig:cluster_characteristics}
\end{figure*}

\subsection{Analyzing identified clusters}
The unsupervised bias detection tool returns three clusters for the CUB 2014 dataset. The characteristics of these clusters are analyzed below.

\textbf{Cluster size:}
Cluster 1 consists of $56,354$ 
 $(26.3\%) $ students, where cluster 2 consists of $123,203$ $ (57.4\%)$ students and cluster 3 of $35,042$ $(16.3\%)$ students.

\textbf{Testing the difference in bias metric per cluster:}
We find stark differences in the bias metric across the three clusters. In the cluster with the highest bias (cluster 3) as shown in Figure \ref{fig:predicted_class_cluster}, 50\% of the students are classified as ``high risk'' by the risk profiling algorithm, versus 3\% and 19\% respectively in clusters 1 and 2. These differences are statistically significant at the 0.1\% level, as indicated in  Table~\ref{tab:test_2014} in Appendix~\ref{sec:tests}, where we report differences between $\barM(U\backslash U_k)$ and $\barM(U_k)$ for the validation set separated for testing, including the $p$-values. 

\begin{figure*}[t]
  \centering
\begin{subfigure}{.25\textwidth}
  \centering
\includegraphics[scale=0.1875]{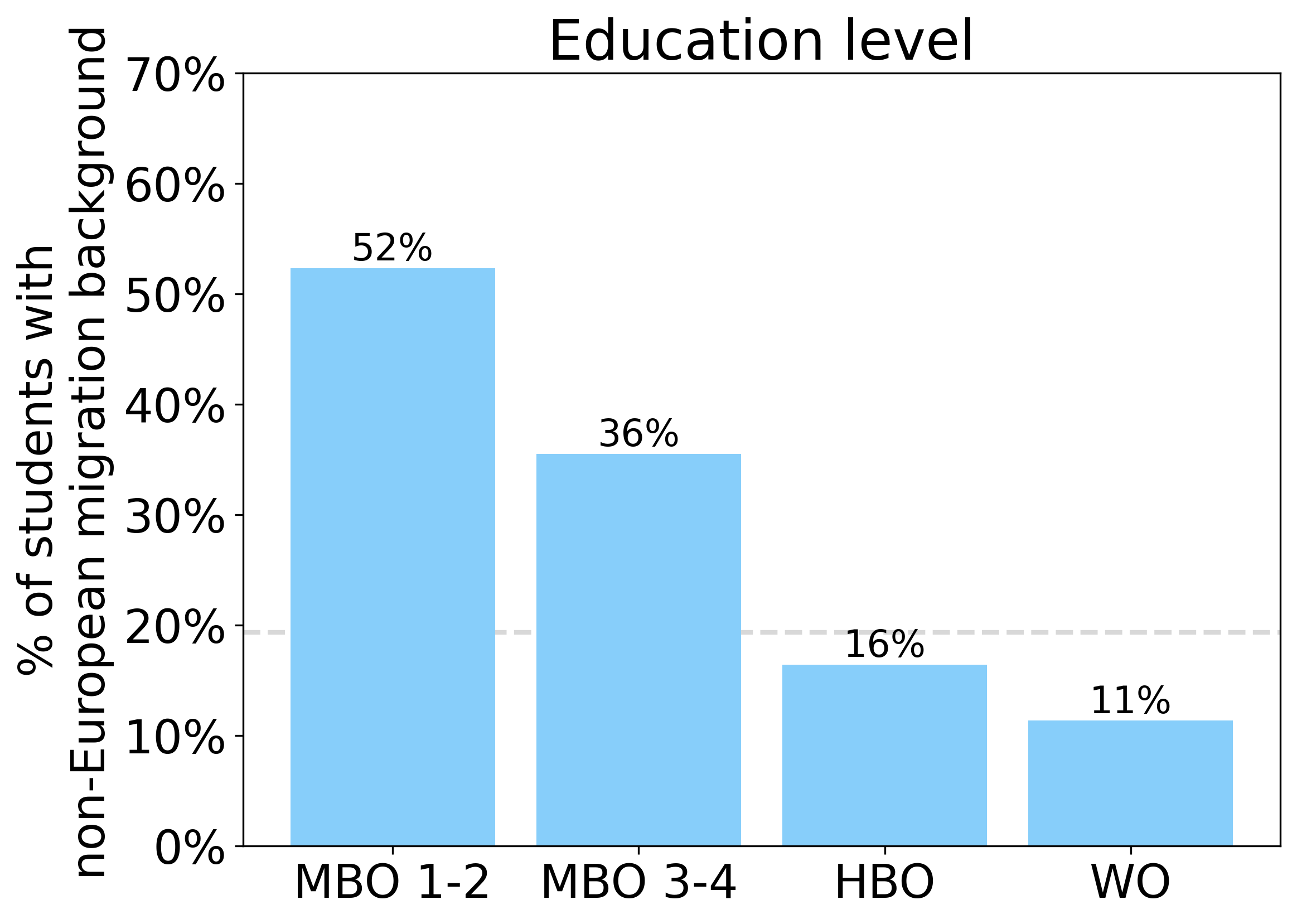}  
  \caption{}
    \label{fig:cluster_educ}

\end{subfigure}
\begin{subfigure}{.3\textwidth}
  \centering
\includegraphics[scale=0.1875]{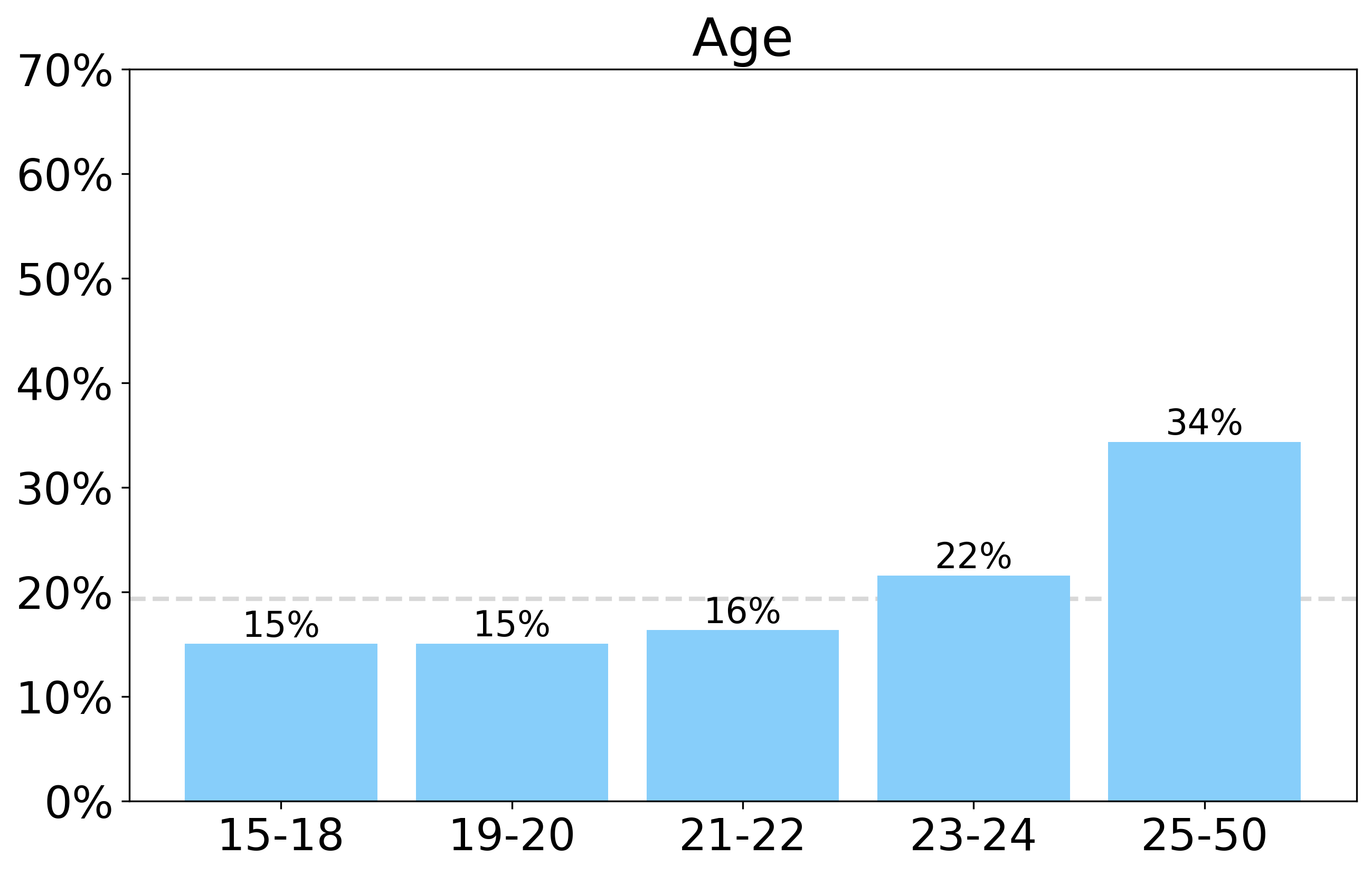} 
  \caption{}
\end{subfigure}
\begin{subfigure}{.44\textwidth}
  \centering
\includegraphics[scale=0.1875]{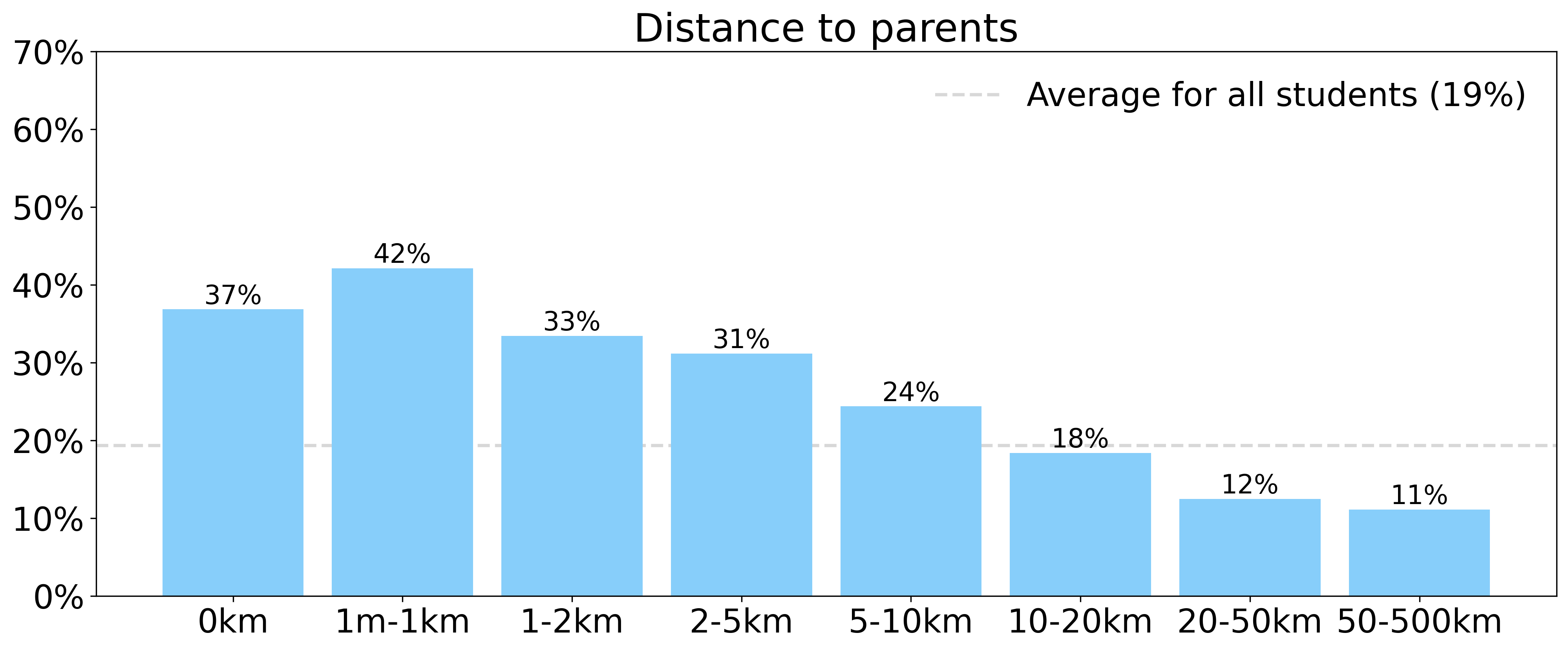}  
  \caption{}
\end{subfigure}
\caption{The percentage of students with a non-European migration background for the three used profiling characteristics: (a) type of education, (b) age and (c) distance to parents for students in the CUB 2014 dataset ($n=248,650$); this number is rounded to the nearest 10 and excludes students where the distance is unknown.}
\label{fig:univariate_proxy}
\end{figure*}

\textbf{Student characteristics per cluster:}
Figure \ref{fig:cluster_characteristics} illustrates the characteristics of students within each cluster. Students in cluster 3, associated with the highest average bias, are more likely to be enrolled in vocational education (MBO 1-2, MBO 3-4), be overrepresented in age groups 15-18 and 25-50, and live on average closer to their parents. In contrast, students in cluster 1, which exhibits the lowest average bias, exclusively attend university, are  more likely to be 19-20 years old, and typically live far from their parents. Students in cluster 2 tend to follow higher professional education (HBO), do not display distinct patterns regarding age and live slightly more often 20-50km away from their parents.

\subsection{Comparing results with aggregate statistics on non-European migration background}
We now examine the extent to which the identified clusters align with the non-European migration background data, using the aggregated data provided by~\citet{CBS}.\footnote{The aggregated statistics are rounded to the nearest ten, hence why there are  $n=248,650$ reported students instead of $248,649$.}
We exclude cases where the distance is unknown, since the algorithm did not apply then.

\textbf{Relationship between clusters and non-European migration background}: As individual-level demographic data on students' origin are unavailable, we estimate the percentage of students with a non-European migration background per cluster using aggregate statistics. This percentage is determined using the weighted average of the proportion of students per cluster, based on the percentage of students with a non-European migration background for each combination\footnote{With 4 types of education, 5 age groups and 8 distance categories, there are in total 160 possible  combinations. Each possible combination is assigned to a unique cluster, e.g. if two students belong to the same combination of education level, age, and distance to parents, they belong to the same cluster.} of the three characteristics~\citep{CBS}.
There are stark differences in the  proportion of students with a non-European migration background across clusters. Figure \ref{fig:non_eu_mig_back_cluster} shows that it is estimated that 41\% of students in cluster 3 have a non-European migration background, compared to 17\% in cluster 2 and 10\% in cluster 1. These results indicate that clusters found by the unsupervised bias detection tool are related to the demographic group of interest. 

\textbf{Proxies for non-European migration background}:
To shed light on the relationship between clusters and non-European migration background, we investigate how characteristics of the clusters (and the risk profiling algorithm) serve as proxies for non-European migration background. This is illustrated in Figure \ref{fig:univariate_proxy}. Students who are enrolled in vocational education (MBO 1-2, MBO 3-4), in the age groups 23-24 and 25-50, and living less than 10km away from their parents on average are more likely to have a non-European migration background, corresponding to the characteristics of students in cluster 3. In contrast, university students (WO), young students, and students living far from their parents are less likely to have a non-European migration background. These characteristics correspond to students in cluster 1; this also aligns with the risk profile logic: younger students and those living closer to their parents are assigned higher risk scores \citep{DutchParliament2024D07565}.

\begin{figure*}[t]
  \centering
\begin{subfigure}{.25\textwidth}
  \centering
\includegraphics[scale=0.25]{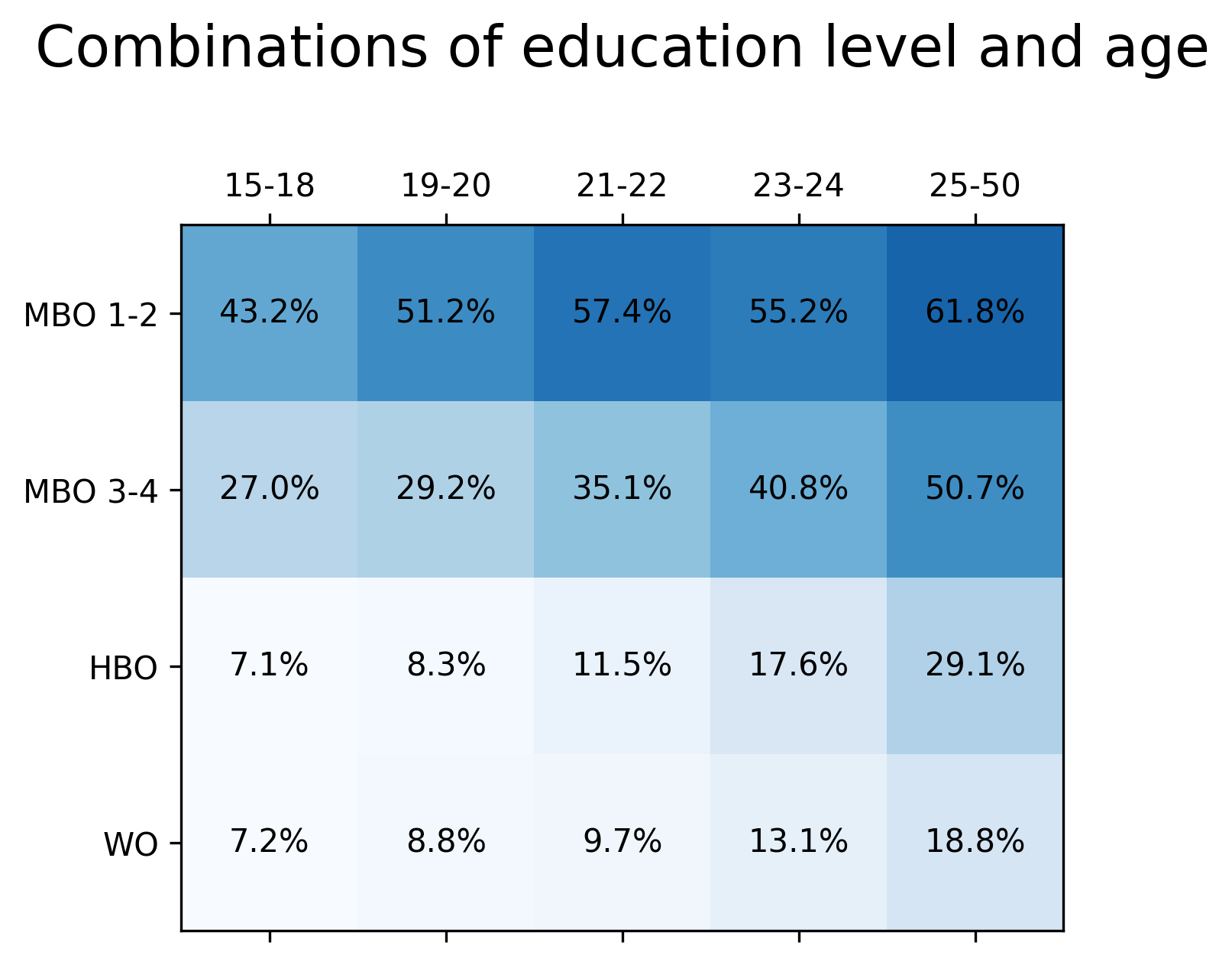}  
  \caption{}
  \label{fig:heatmap_education_age}
\end{subfigure}
\begin{subfigure}{.35\textwidth}
  \centering
\includegraphics[scale=0.25]{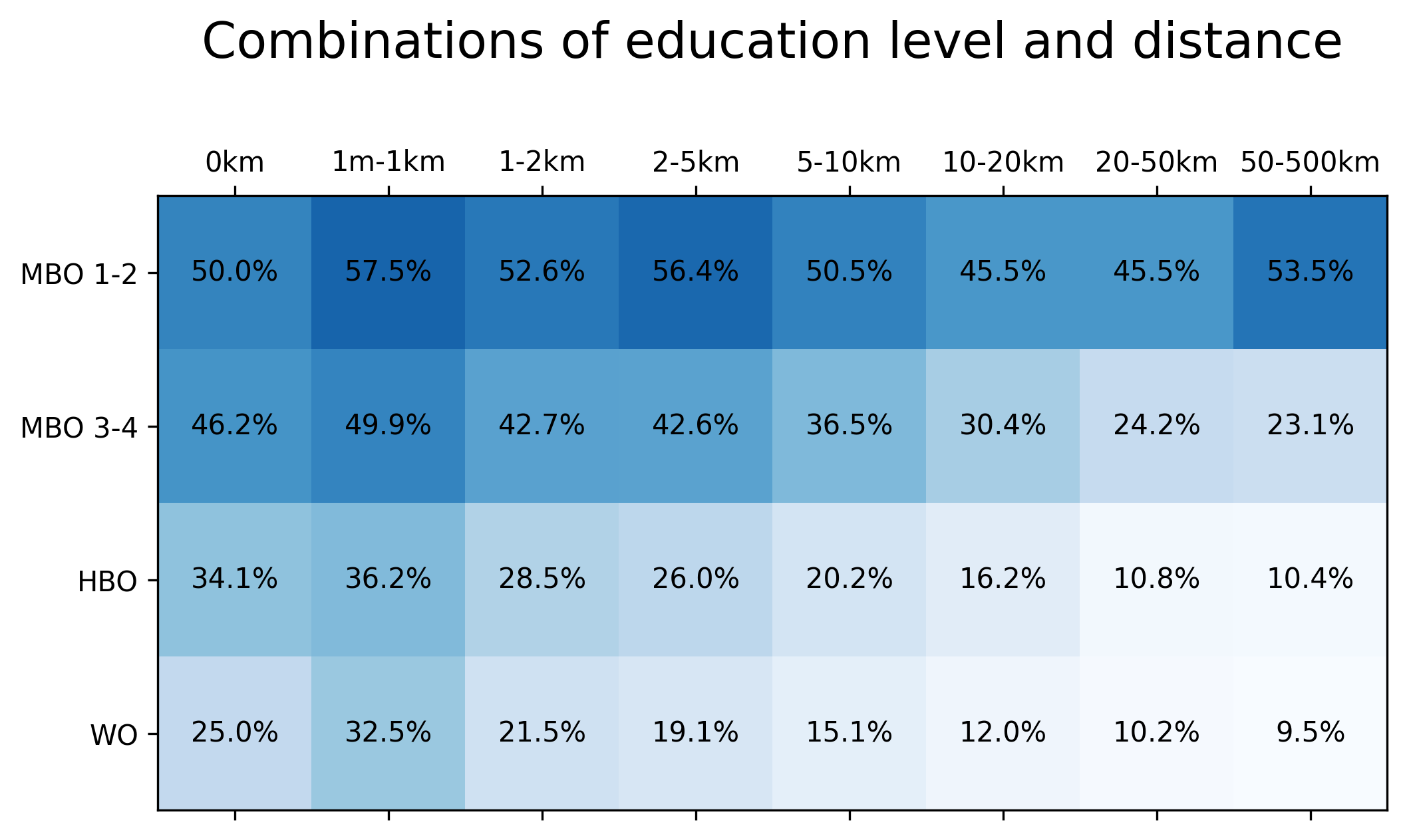} 
  \caption{}
  \label{fig:heatmap_education_distance}
\end{subfigure}
\begin{subfigure}{.35\textwidth}
  \centering
\includegraphics[scale=0.25]{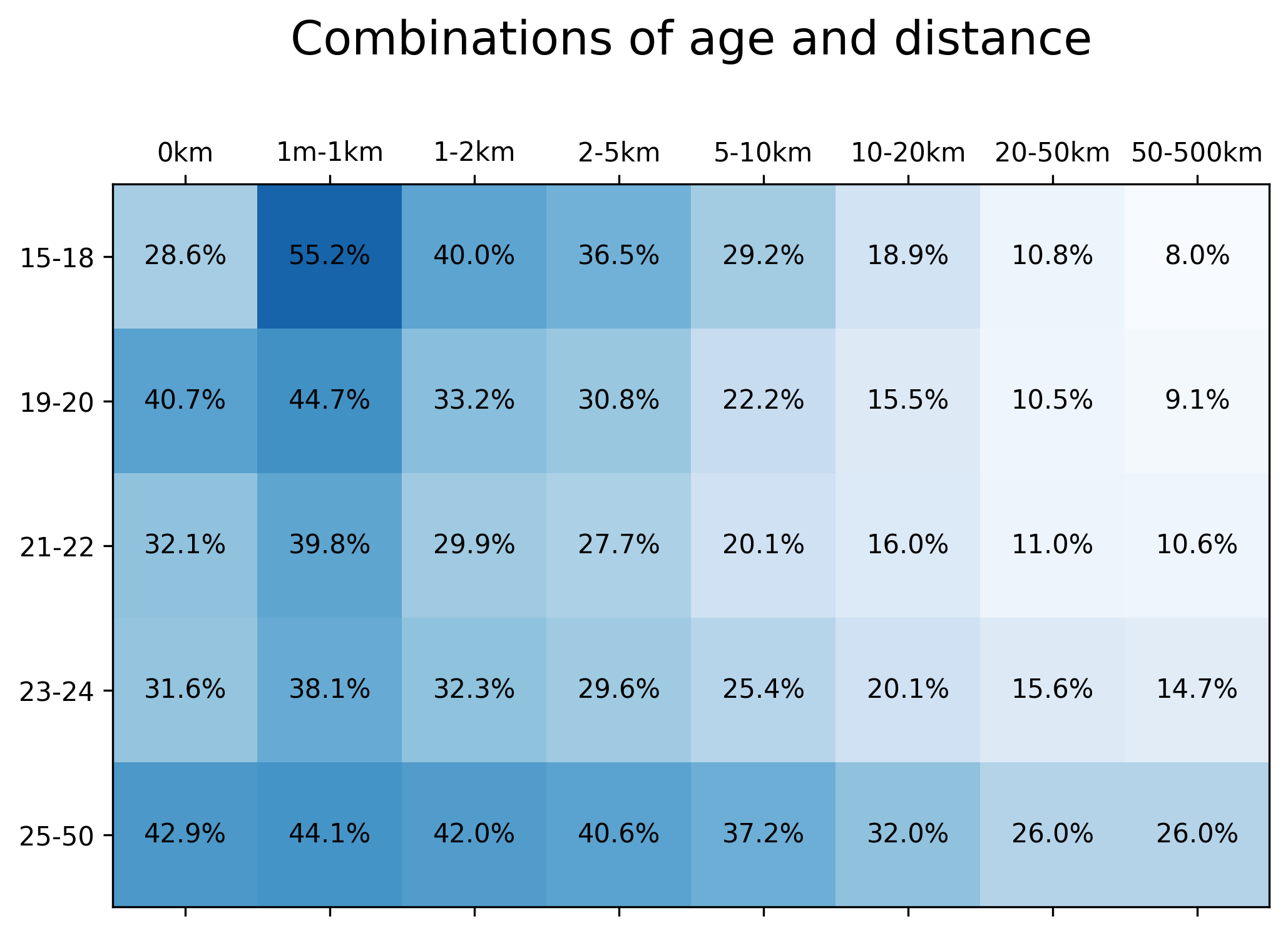}  
  \caption{}
  \label{fig:heatmap_age_distance}
\end{subfigure}
\caption{The percentage of students with a non-European migration background for three (bivariate) combinations of the used profiling characteristics: (a) type of education and age, (b) type of education and distance to parents, and (c) age and distance to parents for the CUB 2014 dataset ($n=248,650$), excluding students where the distance is unknown.}
\label{fig:bivariate_proxy}
\end{figure*}

\textbf{Combinations of profiling characteristics as proxies for non-European migration background}:
In addition to examining single characteristic relationships, we also investigate how specific combinations of characteristics act as a proxy for having a non-European migration background. 
Figure \ref{fig:heatmap_education_age} highlights that young students in university education (WO) are less likely to have a non-European migration background. This result aligns with the characteristics of cluster 1, which consists exclusively of university (WO) students and contains relatively more students aged 19-20. Relatively older students (older than 23) who follow vocational education (MBO 1-2, MBO 3-4) are also more likely to have a non-European migration background. 

Figure \ref{fig:heatmap_education_distance} indicates that students who follow vocational education (MBO) and live relatively close to their parents are more likely to have a non-European migration background. This pattern is reflected in cluster 1, which contains MBO 1-2 and MBO 3-4 students, who tend to live closer to their parents. 
Figure \ref{fig:heatmap_age_distance} shows a strong overrepresentation of students with a non-European migration background in the 15-18 age group who live 1m-1km from their parents. This pattern is somewhat reflected in cluster 3, which contains a higher proportion of 15-18 year old students and students living 1m-1km away from their parents. 

\subsection{Limitations}\label{sec:disc}

\textbf{Limitations of clustering to audit algorithmic systems:}
The unsupervised bias detection tool alone cannot establish prohibited discrimination, as human expertise is essential to evaluate bias within the context of relevant legislation. However, the tool can serve as a starting point for such deliberative assessments. For instance, the clusters that the tool identifies could prompt further investigation into how the characteristics of individuals relate to demographic groups.

\textbf{Limitations of applying the unsupervised bias detection tool to the DUO use case:} 
There are several challenges associated with using aggregated data to evaluate the proportion of students with a non-European migration background in the clusters identified by the tool. For example, this approach requires estimating the percentage of students with a non-European migration background and prevents precise analysis of the joint relationship between the bias metric, clusters and demographic groups. However, this is an unavoidable limitation due to the sensitive nature of the data. %

\section{Conclusion}\label{sec:conc}
As demonstrated through the DUO case study, the unsupervised bias detection tool identifies clusters that correspond to groups of students found to be unequally treated by a risk profiling algorithm. The clustering results highlight proxy characteristics associated with a non-European migration background: vocational education (MBO) and living close to parents, as examined through aggregate statistics provided by Statistics Netherlands.
By improving unsupervised learning methods which are behind the tool we provide, we enhance public knowledge on auditing algorithmic decision-making processes for bias, in the absence of demographic data.
In addition, the unsupervised bias detection method has been implemented as an open-source tool in the form of a Python package, accompanied by a graphical interface in a web application to encourage adoption by a non-technical audience.
The tool provides a scalable method that helps private entities, as well as internal and external auditors, to obtain inferences about seemingly neutral rules or practices that might embed unwanted biases. 
We hope that the contributions from this paper will serve the community by assisting in the detection of biases in algorithmic decision-making systems.

\begin{ack}
For the initial research for this project, Floris Holstege, Mackenzie Jorgensen, Kirtan Padh, and Krsto Prorokovi\'{c} worked under a six month-long funded contract at Algorithm Audit, supported by a SIDN Fund grant for the development of digital commons (Digitale gemeenschapsgoederen). The work of Mackenzie Jorgensen was also supported by the U.K. Research and Innovation under Grant EP/S023356/1 in the UKRI Centre for Doctoral Training in Safe and Trusted Artificial Intelligence (for the majority of the time she worked on this project) and by the Engineering and Physical Sciences Research Council [grant number EP/Y009800/1], through funding from Responsible Ai UK (KP0003). \looseness -1

\end{ack}

\bibliographystyle{ACM-Reference-Format}
\bibliography{refs}

\appendix

\clearpage

\section{The hierarchical bias-aware clustering (HBAC) algorithm}
\label{sec:appendix-hbac-algo}

We provide pseudocode for the HBAC algorithm below and use notation as defined in Section \ref{sec:subsec-notation}. 

\begin{algorithm}[h]
\begin{small}
\KwIn{A dataset $\mathcal{X} = \{ x_1, \ldots, x_N\}$ and bias metric $\{m_1, \ldots m_n\}$. Set the \textit{max\_iterations} and a minimum of samples per cluster $n_{\mathrm{min}}$. }
\KwOut{A partition $\{ \mathcal{C}_1,  \ldots \mathcal{C}_k\}$}
Define the partition = $\{\mathcal{X}\}$ \\
\For{$i \gets 1$ \KwTo max\_iterations} {
    Set $\mathcal{C}$ to be the cluster in partition with the highest standard deviation of metric $M$ among those that have not been selected in any previous iteration. \\
    Split $\mathcal{C}$ into two clusters $\mathcal{C}^{'}$ and $\mathcal{C}^{''}$ (e.g. via $k$-means or $k$-modes) \\
    \If{$\max(\barM(\mathcal{C}^{'}), \barM(\mathcal{C}^{''})) \geq \barM(\mathcal{C}) \littlespace \land \littlespace |\mathcal{C}^{'}| \geq n_{\mathrm{min}} \littlespace \land \littlespace |\mathcal{C}^{''}| \geq n_{\mathrm{min}}$  }{
        Remove $\mathcal{C}$ from partition \\
        Add $\mathcal{C}^{'}$ and $\mathcal{C}^{''}$ to partition
    }
}
\end{small}
\caption{Hierarchical Bias-Aware Clustering}
\label{alg:hbac}
\end{algorithm}

\section{Risk profile}\label{sec:Appendix_Risk_profile}
\begin{figure*}[th]
    \centering
\includegraphics[scale=0.7]{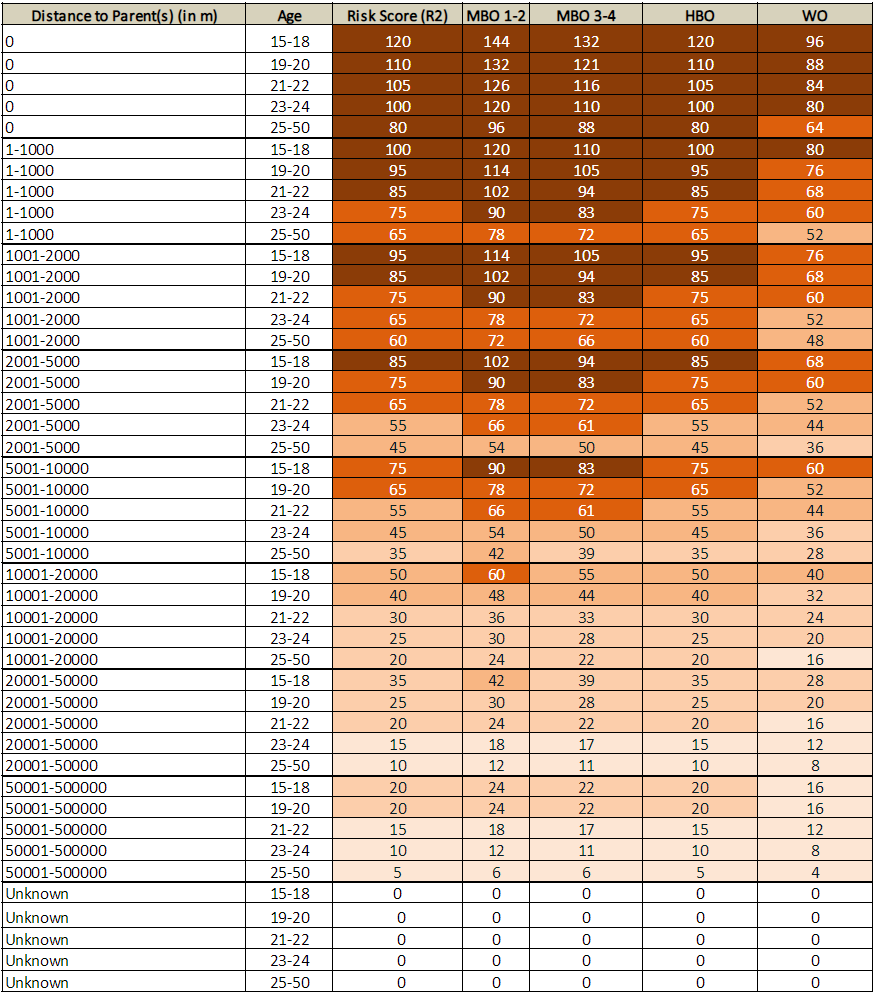}
    \caption{The table of $\mathrm{R}_2$ risk factor determined by the combination of the distance category and the age category of the student.}
    \label{fig:R2}
\end{figure*}

\begin{figure}[t]
    \centering
       \includegraphics[scale=0.4]{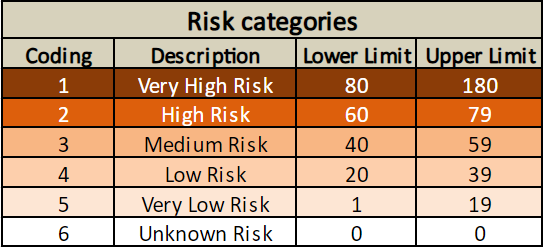}
        \caption{A tabular representation of the risk categories and the corresponding lower and upper limit of the risk score. These limits indicate when a student is deemed to be ``high risk'' for improper allocation of a college grant, i.e., if the risk score for a student is determined between $80$ and $180$, students were allocated to risk category 1.}
    \label{fig:risk_cat}
\end{figure}

The risk score for student $u_{1 \leq i \leq N}$ is determined as follows: $\text{risk score}_{u_i} = \mathrm{R}_1 \times (\mathrm{R}_2 + \mathrm{R}_3)$. 
$\mathrm{R}_1$ is the risk factor assigned to different types of education (see Table \ref{tab:riskfactor_1}). $\mathrm{R}_2$ is a risk factor determined by the combination of the distance category and the age category of the student (see Figure \ref{fig:R2}). $\mathrm{R}_3$ is an adjustment based on whether there were particular deviations between the age of the student known to DUO and the age known to the Municipal Basic Administration (GBA), combined with information about how long the student has been living away from home (see Table \ref{tab:r3}). Based on these risk factors, a risk score is computed which are binned in 6 risk categories (see Figure \ref{fig:risk_cat}). 

\begin{table*}[th]
\centering
\begin{tabular}{l c}
    \toprule
    Type of education & Factor \\
    \midrule
    MBO 1-2        & 1.2                                                         \\
    MBO 3-4        & 1.1                                                         \\
    HBO            & 1.0                                                         \\
    WO             & 0.8   \\                   
    \bottomrule
    \end{tabular}
    \caption{The risk factor $\mathrm{R}_1$  for the different types of education.}
\label{tab:riskfactor_1}
\end{table*}

\begin{table*}[t]
\centering
\footnotesize
\begin{tabular}{llllllc}
\toprule
\textbf{Curr. Age}  & \textbf{Curr. Age} & \textbf{Age Reg.} & \textbf{Age Reg.}  & \textbf{Curr. Age (GBA)}  & \textbf{Curr. Age (GBA)} & \textbf{Risk Factor} \\
From & Up Until & From & Up Until & From  &  Up Until &  $\mathrm{R}_3$ \\
\midrule
21                 & 22               & 17                      & 18                    & 17                      & 18                    & 5              \\
21                 & 22               & 17                      & 18                    & 19                      & 20                    & 0              \\
21                 & 22               & 19                      & 20                    & 19                      & 20                    & 0              \\
23                 & 24               & 17                      & 18                    & 17                      & 18                    & 15             \\
23                 & 24               & 17                      & 18                    & 19                      & 20                    & 10             \\
23                 & 24               & 17                      & 18                    & 21                      & 22                    & 0              \\
25                 & 65               & 17                      & 18                    & 17                      & 18                    & 30             \\
25                 & 65               & 17                      & 18                    & 19                      & 20                    & 25             \\
25                 & 65               & 17                      & 18                    & 21                      & 22                    & 15             \\
25                 & 65               & 17                      & 18                    & 23                      & 24                    & 0              \\
25                 & 65               & 17                      & 18                    & 25                      & 65                    & 0              \\
25                 & 65               & 19                      & 20                    & 19                      & 20                    & 25             \\
25                 & 65               & 19                      & 20                    & 21                      & 22                    & 0              \\
25                 & 65               & 19                      & 20                    & 23                      & 24                    & 0              \\
25                 & 65               & 19                      & 20                    & 25                      & 65                    & 0              \\
25                 & 65               & 21                      & 22                    & 21                      & 22                    & 15             \\
25                 & 65               & 21                      & 22                    & 23                      & 24                    & 0              \\
25                 & 65               & 23                      & 24                    & 23                      & 24                    & 0 \\
\bottomrule
\end{tabular}
\caption{Values of the $\mathrm{R}_3$ risk factor depend on the current age (`Curr. Age') of students (referring to the current age of a student as officially registered by DUO), the age when registered (`Age Reg.', referring to the age known by DUO when the student was registered at an address different from their parental address), and the current age (`Current age (GBA)', referring to the age known by the  Municipal Basic Administration (GBA)).}
\label{tab:r3}
\end{table*}

\clearpage

\section{Applying the unsupervised bias detection tool to the CUB 2019 dataset}\label{sec:CUB-2019}

We replicate all the figures presented in Section \ref{sec:2014-results} for the student population of 2019, i.e., the CUB 2019 dataset ($n=50,233$). This student population differs from the student population in 2014 in a key aspect. After 2015, students no longer received the grant for living outside of their parents home if they were following higher professional (HBO) or university (WO) education. The HBO and WO students included in the CUB 2019 dataset started receiving the college grant before 2015. Because of this difference, we separately analyze the CUB 2019 and 2014 dataset. 

We follow the exact same procedure as outlined in Section \ref{sec:DUO}, only changing the options for the minimum number of samples to $1,000, 2,000, 3,500, \text{and } 5,000$. We make this change because the student population of 2019 is smaller than the one in 2014. 

The clustering results from the 2019 data are broadly in line with the results from 2014. One difference is that we find 2 instead of 3 clusters. This can be explained by the fact that there are much less HBO and WO students present in this student population. 

\begin{figure*}[th]
 \centering
\begin{subfigure}{.25\textwidth}
  \centering
\includegraphics[scale=0.195]{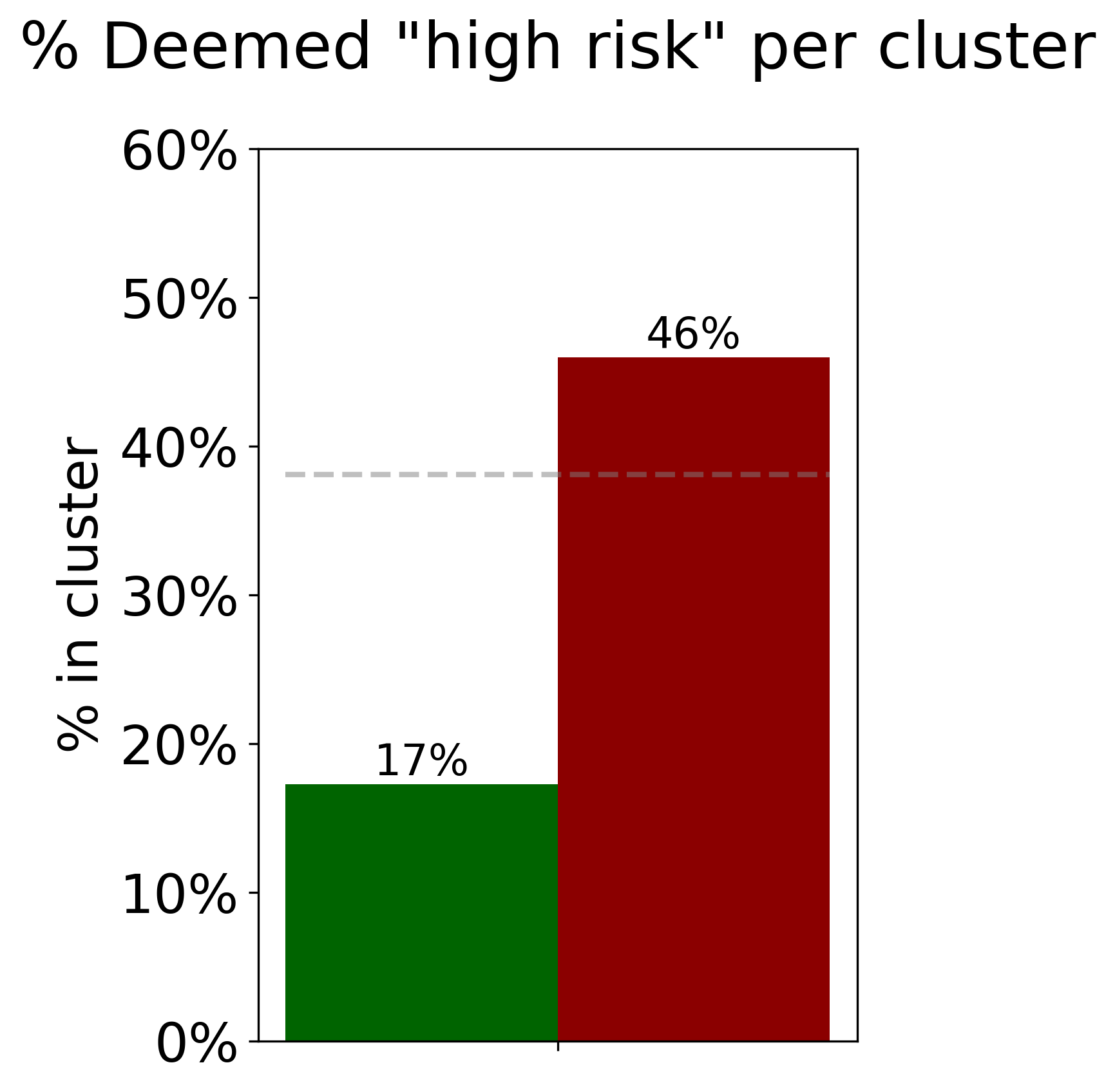}  
\caption{}
\label{fig:predicted_class_cluster_19}

\end{subfigure}
\begin{subfigure}{.5\textwidth}
  \centering
\includegraphics[scale=0.195]{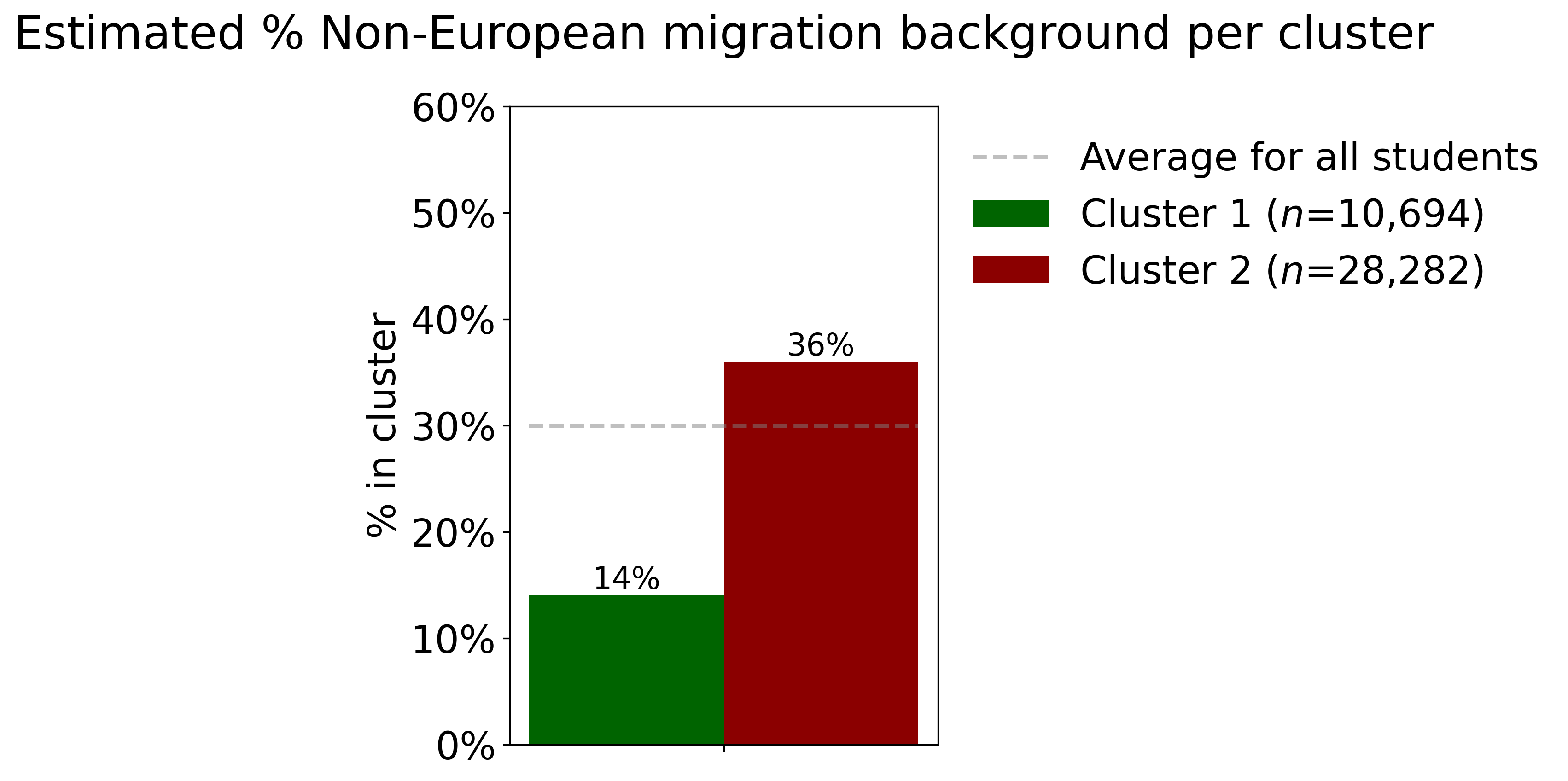}  
  \caption{}
    \label{fig:non_eu_mig_back_cluster_19}

\end{subfigure}
\caption{The percentage of students (a) deemed ``high risk'' and (b) non-European migration background for identified clusters based on the CUB 2019 dataset, excluding students for which the risk profile was unknown  ($n=38,976$).}
\label{fig:bivariate_bias_background_19}
\end{figure*}

Again, there are stark differences in the bias metric between clusters. For instance, Figure \ref{fig:predicted_class_cluster_19} shows that in cluster 2, 46\% of the students are deemed "high risk", whereas only 17\% are deemed "high risk" in cluster 1. This difference is statistically significant at the 0.1\% level, as indicated in Table \ref{tab:test_2019} in Appendix \ref{sec:tests}.

We also (again) observe stark differences per cluster in the estimated percentage of students with a non-European migration background. In Figure \ref{fig:non_eu_mig_back_cluster_19} it is shown that the estimated percentage of students with a non-European migration background in cluster 2 is 36\%, where it is only 14\% in cluster 1. 

\begin{figure*}[th]
  \centering
\begin{subfigure}{.25\textwidth}
  \centering
\includegraphics[scale=0.1875]{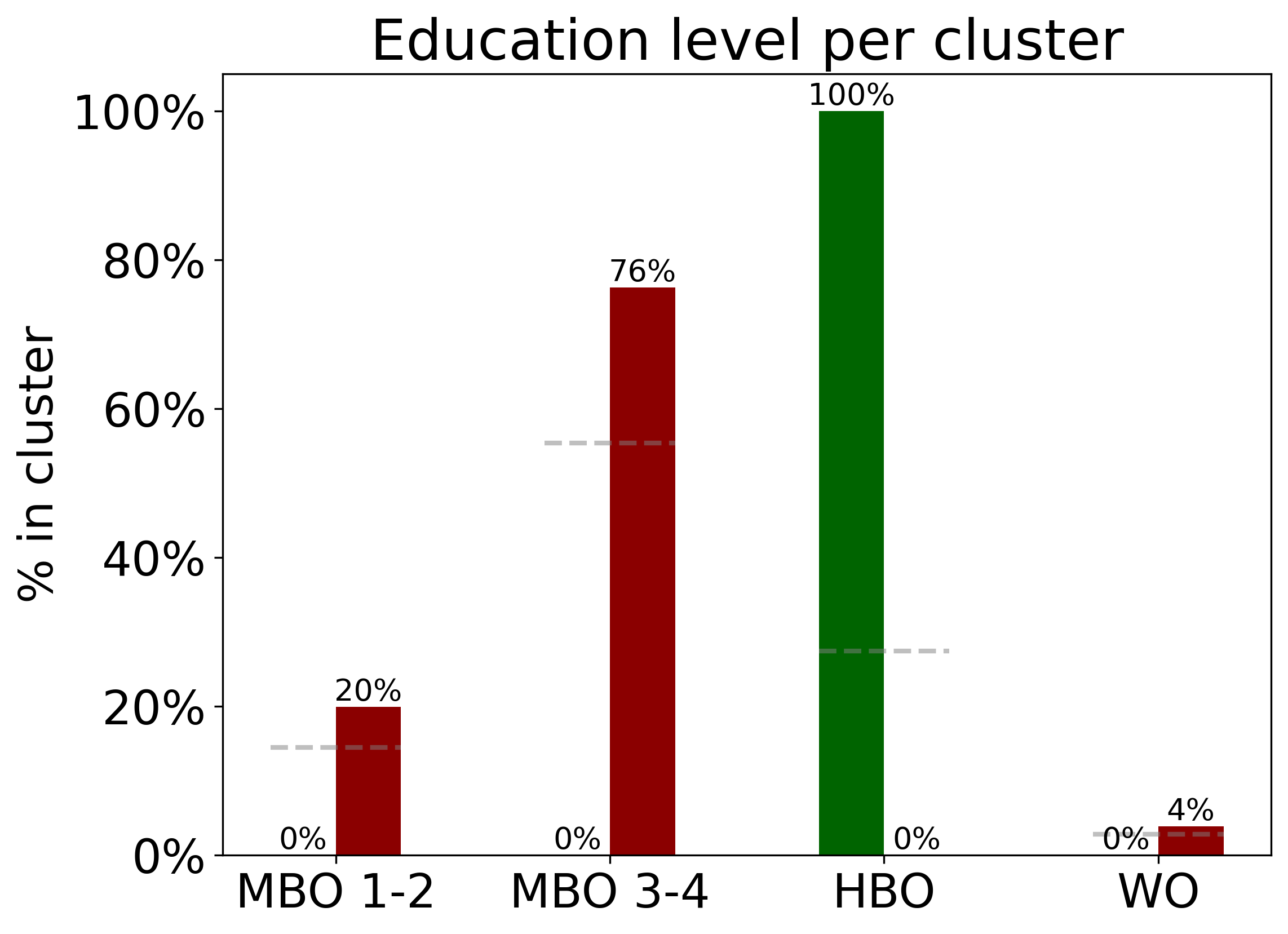}  
  \label{fig:cluster_educ_19}
  \caption{}
\end{subfigure}
\begin{subfigure}{.3\textwidth}
  \centering
\includegraphics[scale=0.1875]{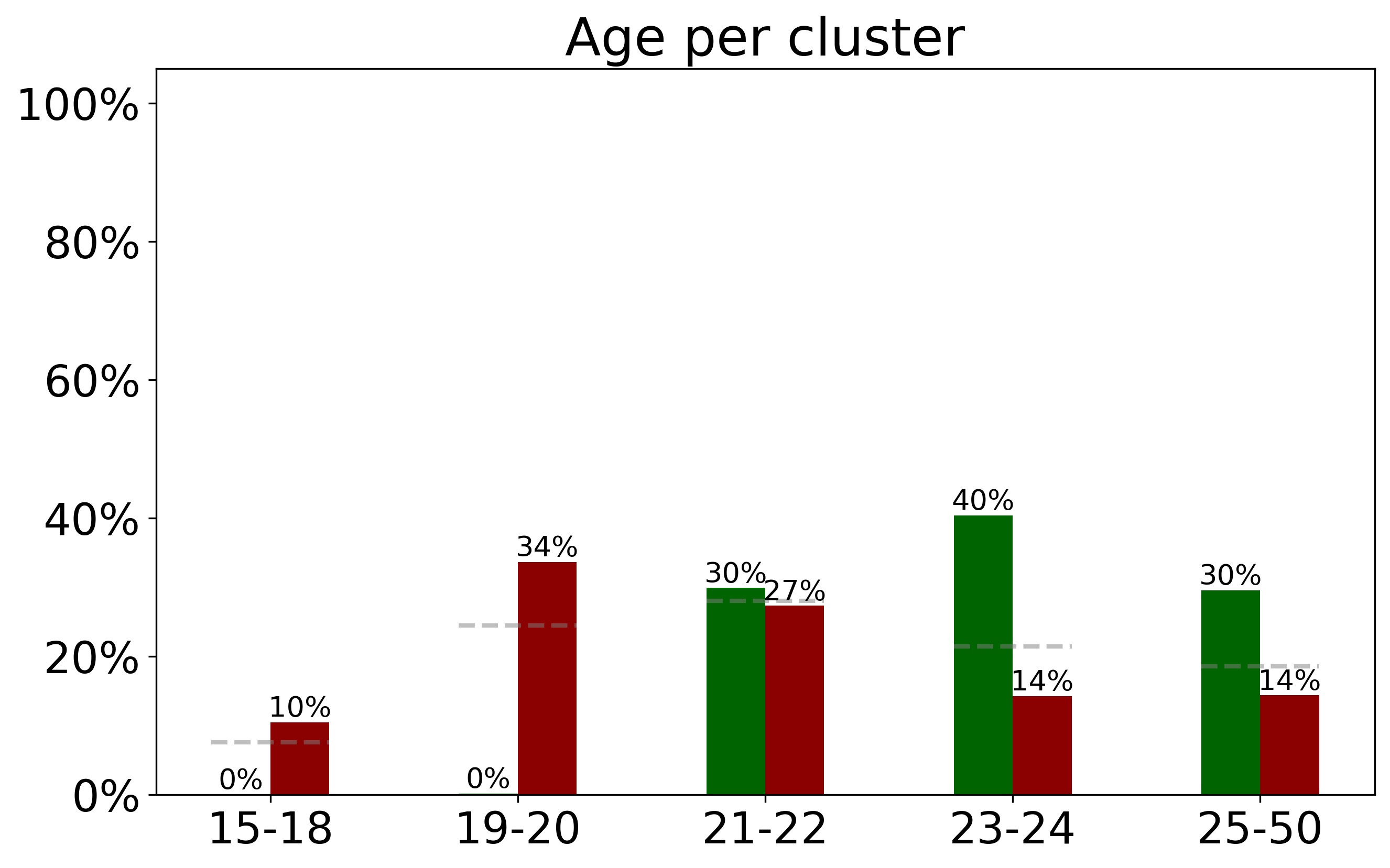}  
  \caption{}
\end{subfigure}
\begin{subfigure}{.44\textwidth}
  \centering
\includegraphics[scale=0.1875]{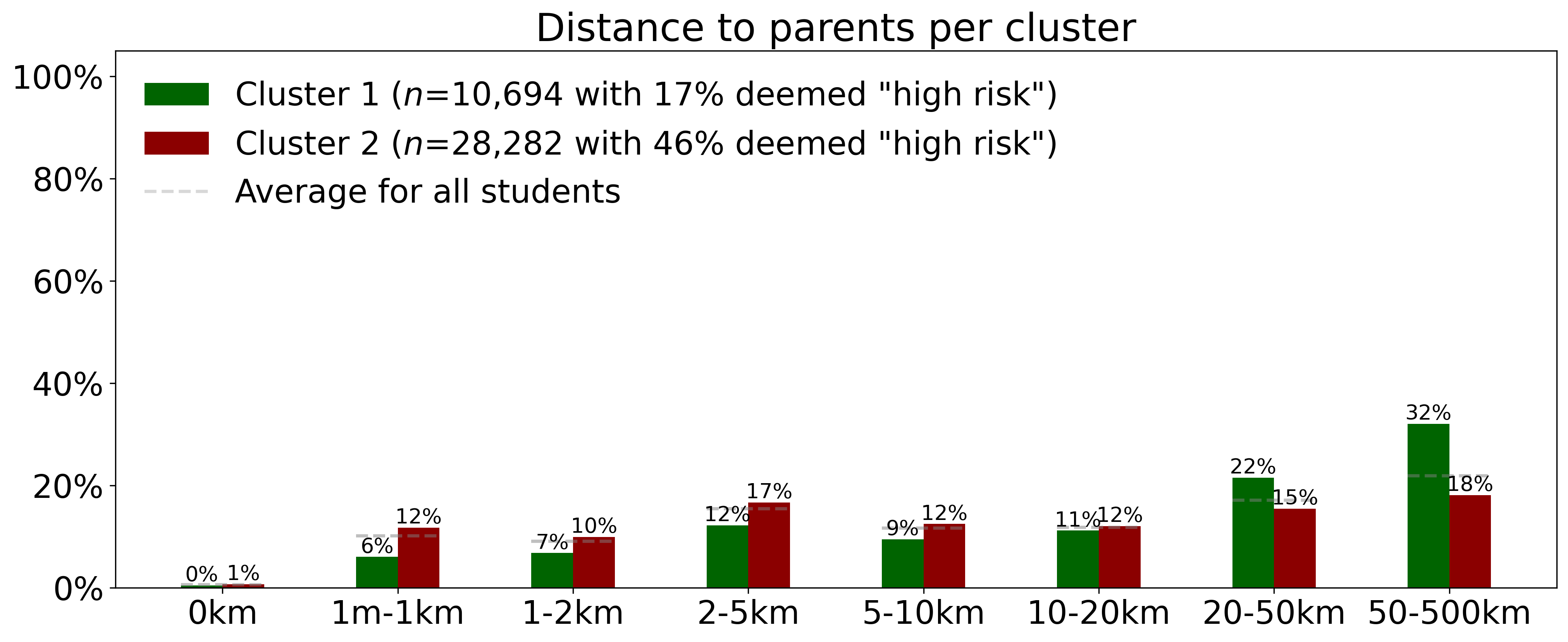}  
  \label{fig:cluster_distance_19}
  \caption{}
\end{subfigure}
\caption{The percentage of students in each cluster for the three characteristics used in the risk profiling algorithm: (a) type of education, (b) age and (c) distance to parents within the CUB 2019 dataset, excluding students for which the risk profile was unknown  ($n=38,976$). For each subgroup, the average is indicated by the dashed line.}
\label{fig:cluster_characteristics_2019}
\end{figure*}

Cluster 1 (the cluster with relatively lower bias) consists exclusively of students following higher professional education (HBO), students who live relatively further away from their parents, and relatively older students, i.e., exclusively consisting of students older than 21. Cluster 2 (the cluster with a relatively higher bias) consists of 96\% of students following vocational education (MBO 1-2, MBO 3-4), students who live relatively closer to their parents, and relatively younger students. We include these characteristics found in each cluster as shown in the clustering results in Figure \ref{fig:cluster_characteristics_2019}.

\begin{figure*}[t]
  \centering
\begin{subfigure}{.25\textwidth}
  \centering
\includegraphics[scale=0.1875]{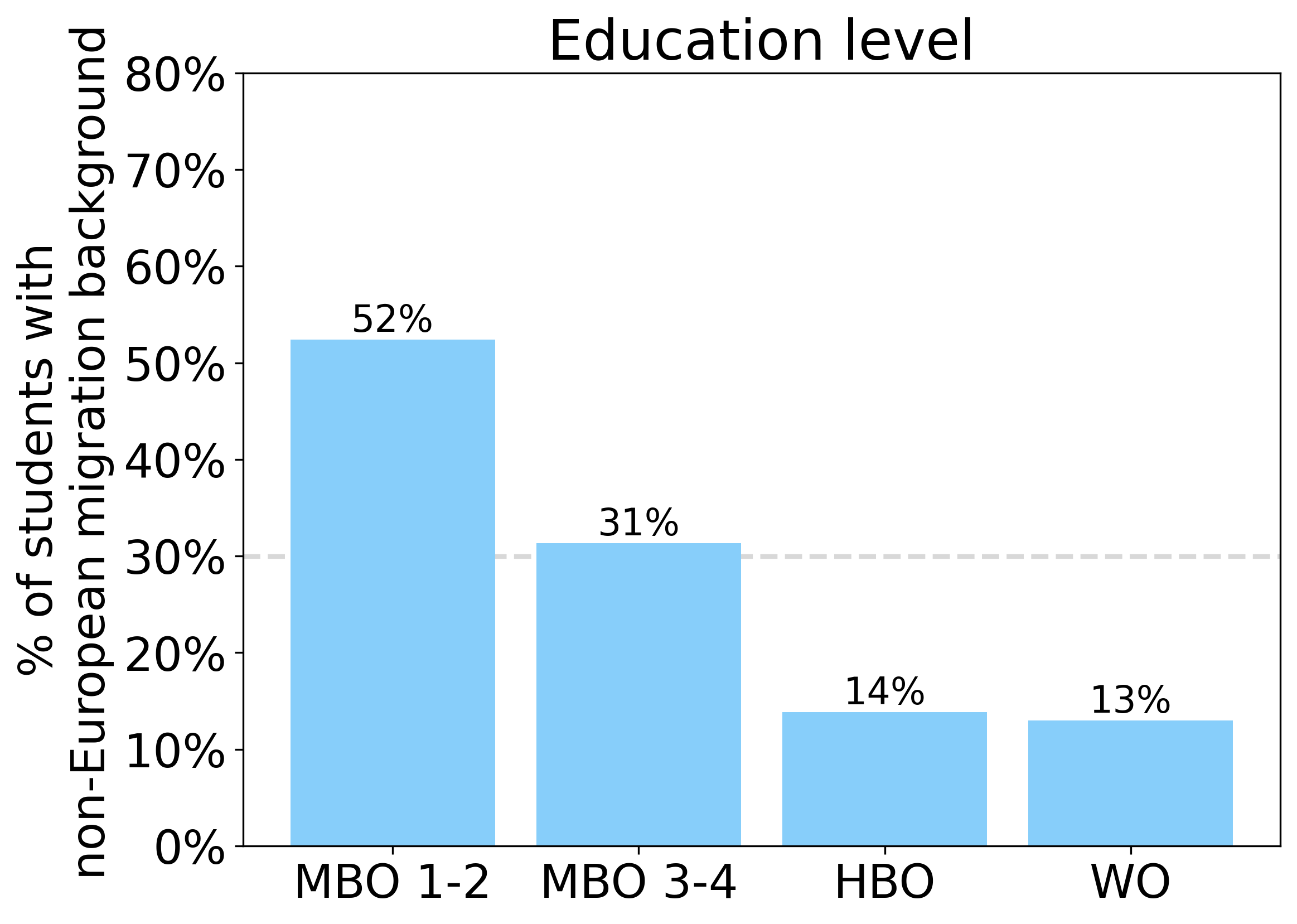}  
  \caption{}
\end{subfigure}
\begin{subfigure}{.3\textwidth}
  \centering
\includegraphics[scale=0.1875]{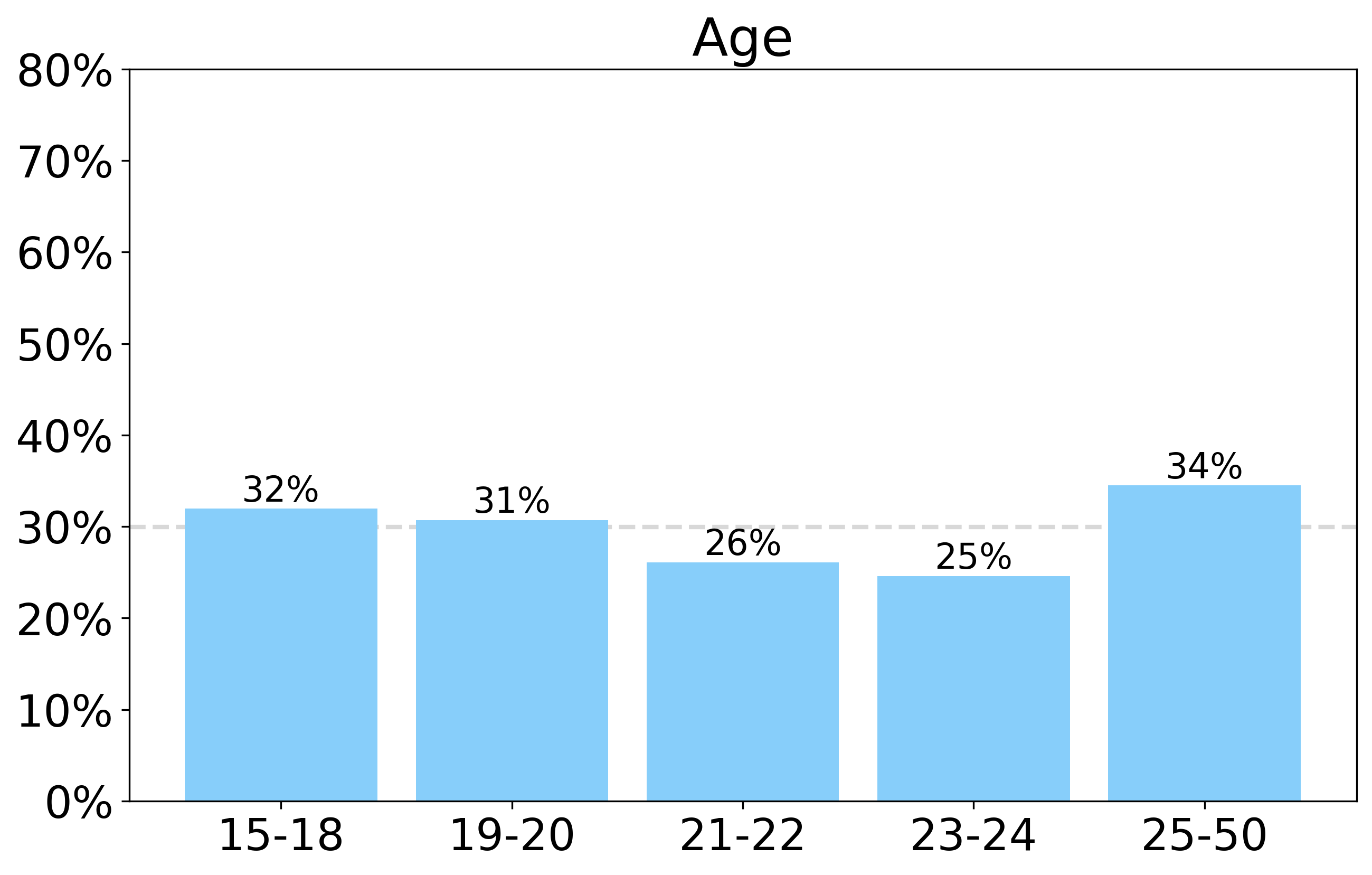} 
  \caption{}
\end{subfigure}
\begin{subfigure}{.44\textwidth}
  \centering
\includegraphics[scale=0.1875]{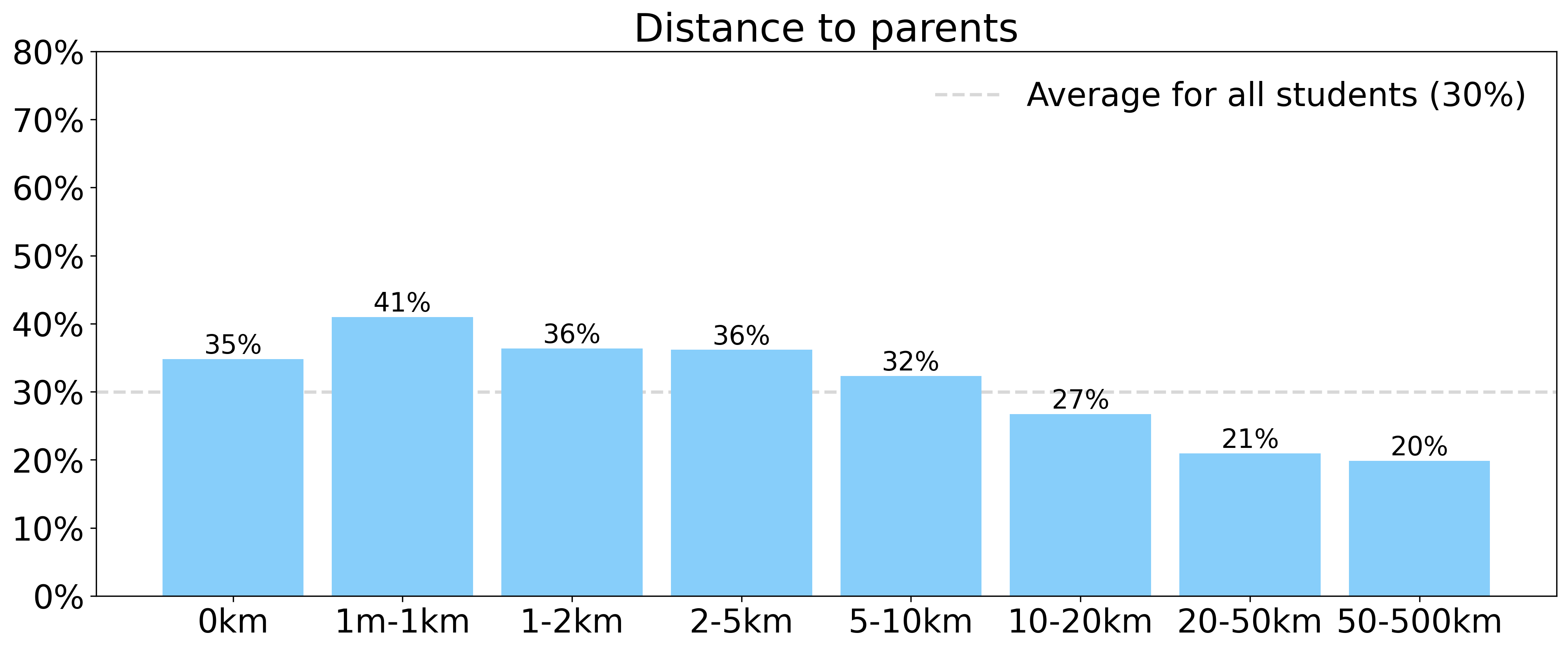}  
  \caption{}
\end{subfigure}
\caption{The percentage of students with a non-European migration background for the three used profiling characteristics: (a) type of education, (b) age and (c) distance to parents for students in the CUB 2019 dataset ($n=50, 230$), excluding students where the distance is unknown.}
\label{fig:univariate_proxy_2019_19}
\end{figure*}

\begin{figure*}[t]
  \centering
\begin{subfigure}{.25\textwidth}
  \centering
\includegraphics[scale=0.25]{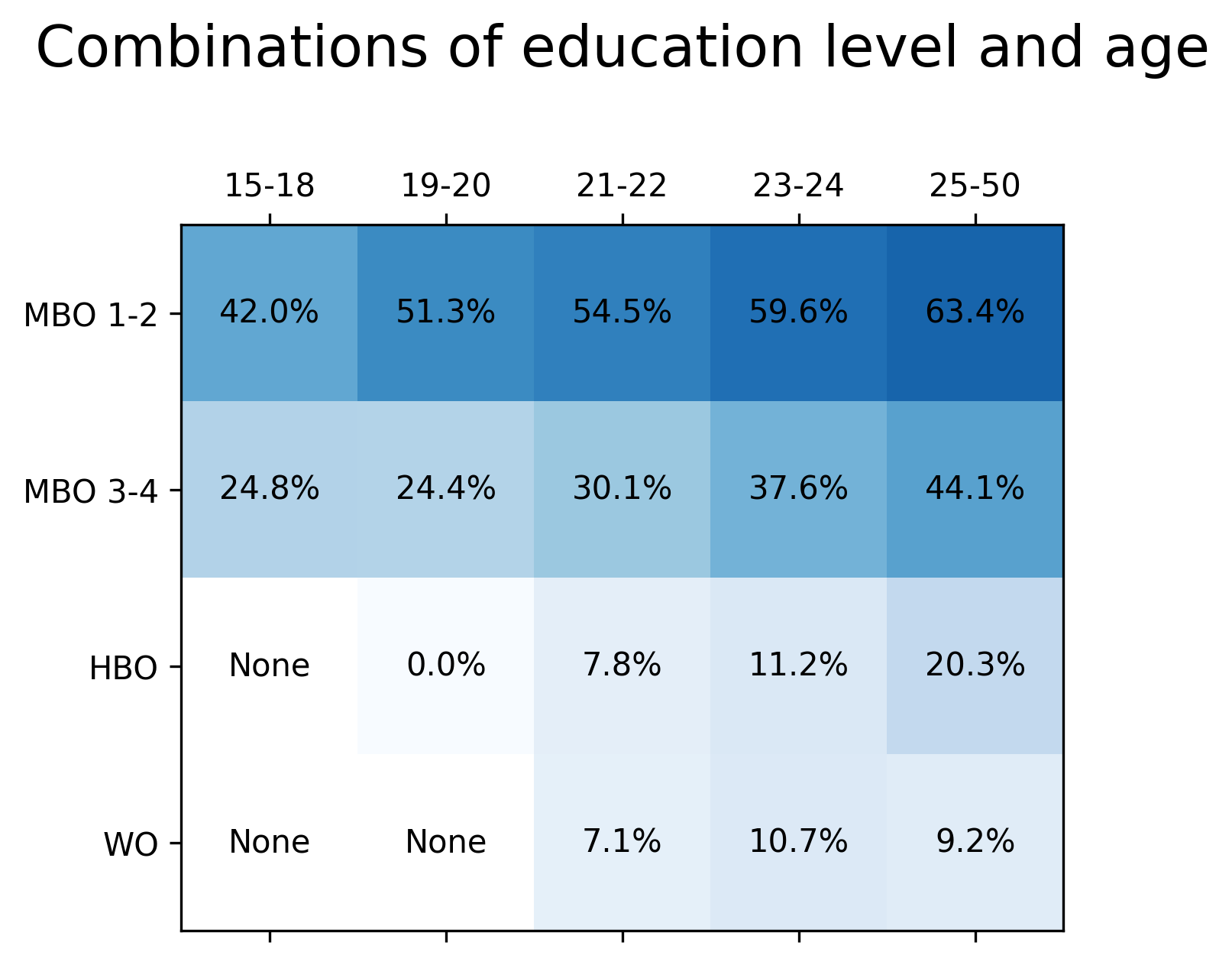}  
  \caption{}
\end{subfigure}
\begin{subfigure}{.35\textwidth}
  \centering
\includegraphics[scale=0.25]{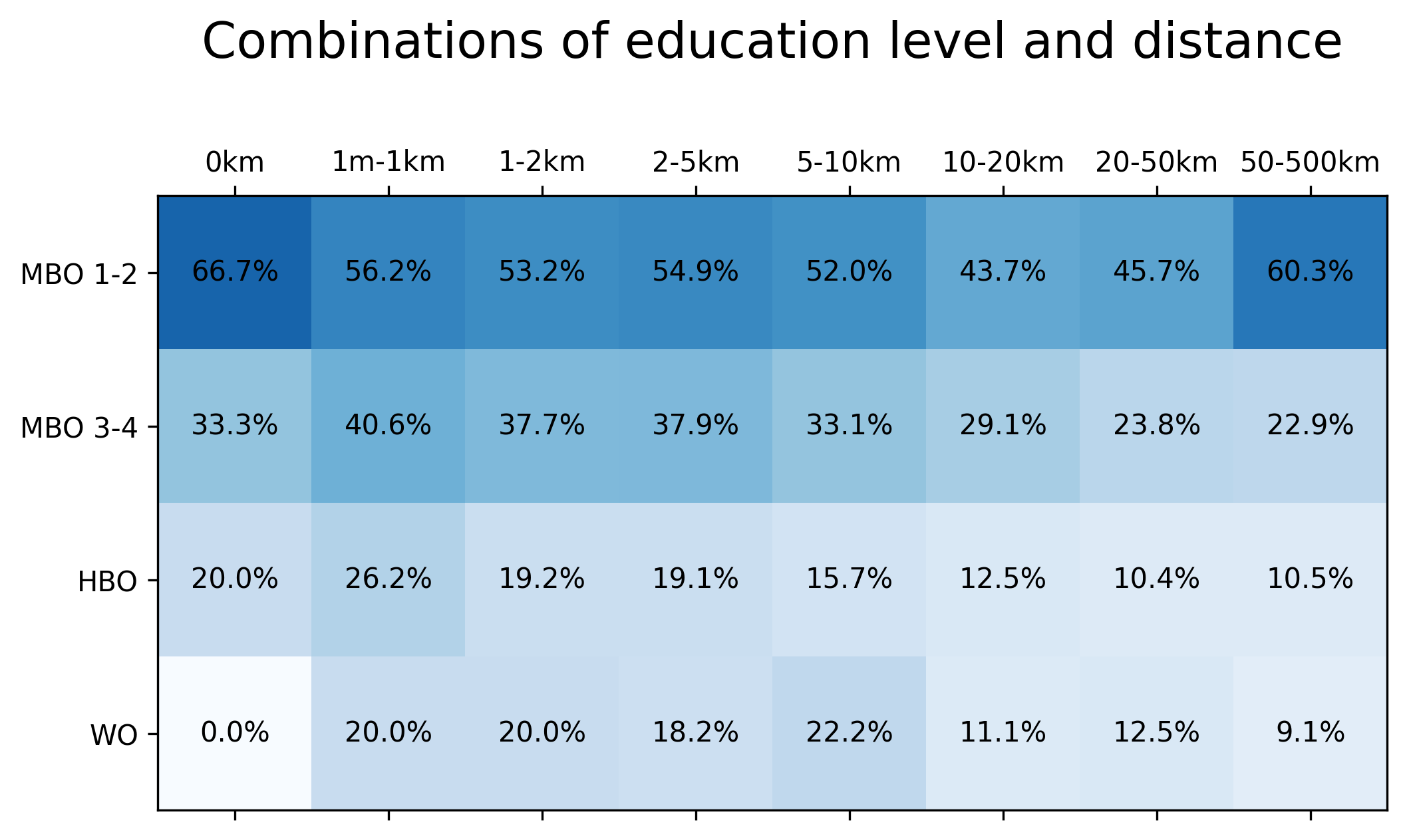} 
  \label{fig:cluster_age_19}
  \caption{}
\end{subfigure}
\begin{subfigure}{.35\textwidth}
  \centering
\includegraphics[scale=0.25]{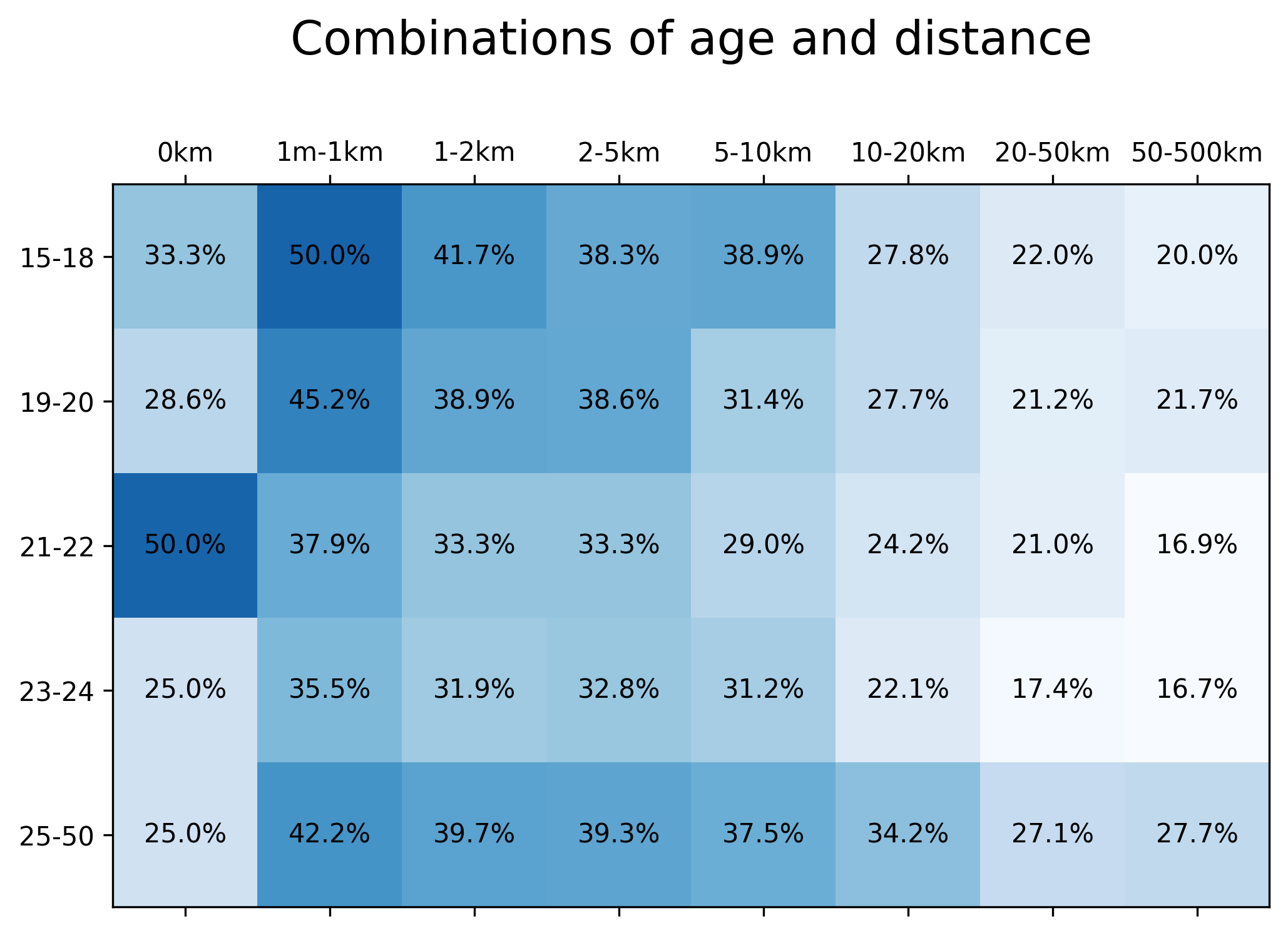}  
  \caption{}
\end{subfigure}
\caption{The percentage of students with a non-European migration background for three (bivariate) combinations of the used profiling characteristics: (a) type of education and age, (b) type of education and distance to parents, and (c) age and distance to parents for the CUB 2019 dataset ($n=50, 230$), excluding students where the distance is unknown. The `None' indicates there were no observations available for these categories.}
\label{fig:bivariate_proxy_2019}
\end{figure*}

Based on the aggregated data available on the origin of students, as provided by Statistics Netherlands ($n=50,230$),\footnote{The aggregated statistics are rounded to the nearest ten; hence, why there are $n=50,230$ reported students instead of ($n=50,233$).} we again observe how particular characteristics used in the risk profile serve as proxies for non-European migration background (see Figure \ref{fig:univariate_proxy_2019_19}). Similar to 2014, students following vocational education (MBO 1-2, MBO 3-4) are relatively more likely to have a non-European migration background, and students living relatively closer to their parents, specifically less than 5km, are also relatively more likely to have a non-European migration background. This corresponds to the characteristics of students in cluster 2. Particular combinations of characteristics also observe as proxies for non-European migration background (see Figure \ref{fig:bivariate_proxy_2019}). For instance, similar to 2014, students older then 23 years who are enrolled in vocational education (MBO 1-2, MBO 3-4) are more likely to have a non-European migration background. 

\section{Statistical test details and results}
\label{sec:tests}

Here, we provide details for the statistical tests  used to investigate whether or not differences between the bias metric per cluster are statistically significant. 
After HBAC clustering, we test the following null versus alternative hypotheses for each of the identified clusters:
\begin{align}
    H_0: \barM(U\backslash U_k)  = \barM(U_k), \littlespace H_A: \barM(U\backslash U_k)  \neq \barM(U_k).
\end{align}
If $\barM(U\backslash U_k) \neq \barM(U_k)$, the bias is found to be higher or lower for a particular cluster found by the clustering algorithm.  To test these hypotheses, we use a two-sample $t$-test when the bias metric is continuous, and Pearson's $\chi^2$-test when the bias metric is a binary variable. For either case, the test is two-tailed, as we allow $\barM(U_k)$ to be both lower and higher than $\barM(U\backslash U_k)$, meaning that we can detect bias both when it is greater or less for a group compared to the rest.

For valid statistical inference, the null and alternative hypotheses must be defined before observing the data. If we conduct our statistical test on data that was also used to identify clusters, this condition is violated.  Since the clusters are selected to maximize the difference in the bias metric, we will therefore tend to incorrectly conclude that there is bias when there truly is none.
The broader issue of testing data-driven hypotheses---often called \emph{post-selection inference}---has been studied extensively in the context of regularized regression~\citep{Kuchibhotla2022}. A long-standing practice in that literature is to use different partitions of the data for model estimation and inference, referred to as \textit{sample splitting} \citep{larson1931shrinkage, stone1974cross}. We apply this technique by fitting the HBAC algorithm on a subset of the data and subsequently inferring the clusters and their bias on a \textit{new} subset of data. Importantly, this set cannot be used to determine our hyperparameters.

When testing for multiple $K > 2$ clusters, an issue is that without adjustments, the likelihood of (incorrectly) detecting a statistically significant difference under the null hypothesis increases. We address this by using a Bonferroni correction \citep{hochberg1987multiple}. 

In Table \ref{tab:test_2014}, we report results for the student population in 2014; in Table \ref{tab:test_2019}, we provide results for the student population in 2019.

\begin{table*}[th]
\centering
\begin{tabular}{l c c c c c}
\toprule
\textbf{Cluster} & 
\textbf{\(n\) in cluster} & 
\textbf{\begin{tabular}{c}High risk (\%)\\ in cluster\end{tabular}} & 
\textbf{\begin{tabular}{c}High risk (\%)\\ outside cluster\end{tabular}} & 
\textbf{Difference (\%)} & 
\textbf{P-value} \\
\midrule
1 & 11,356 & 3.39 & 25.58 & -22.19 & $<10^{-16}$ \\
2 & 24,650 & 18.88 & 20.83 & -1.95  & $1.73\times 10^{-6}$ \\
3 & 6,914  & 49.46 & 13.99 & 35.47  &  $<10^{-16}$ \\
\bottomrule
\end{tabular}
\caption{Results of testing difference in bias metrics across clusters returned by the unsupervised bias detection tool on the validation set for the year 2014 ($n=42,920$), using a Pearson's $\chi^2$-test.  The `$<10^{-16}$' is used to indicate that the $p-$value was smaller than the smallest decimal that can be represented due to integer underflow.}
\label{tab:test_2014}
\end{table*}

\begin{table*}[th]
\centering
\begin{tabular}{l c c c c c}
\toprule
\textbf{Cluster} & 
\textbf{\(n\) in cluster} & 
\textbf{\begin{tabular}{c}High risk (\%)\\ in cluster\end{tabular}} & 
\textbf{\begin{tabular}{c}High risk (\%)\\ outside cluster\end{tabular}} & 
\textbf{Difference (\%)} & 
\textbf{P-value} \\
\midrule
1 & 2,065 & 17,29  & 46,34 & -22.19 & $<10^{-16}$\\ 
\bottomrule
\end{tabular}
\caption{Results of testing difference in bias metrics across clusters returned by the unsupervised bias detection tool on the validation set for the year 2019 ($n=7,796$), using a Pearson's $\chi^2$-test.  The `$<10^{-16}$' is used to indicate that the $p-$value was smaller than the smallest decimal that can be represented due to integer underflow. In this case, we only conduct a single statistical test, since there are only two clusters returned.}
\label{tab:test_2019}
\end{table*}

\clearpage

\section{Simulation study to evaluate the clustering algorithm}
\label{sec:addendum_sim}

This section provides simulations to support several of the design choices made when applying the Hierarchical Bias-Aware Clustering (HBAC) algorithm, and develops empirical guidance for how to best use it. Similarly to \citet{MISZTALRADECKA2021102519}, we study the performance of the HBAC algorithm on synthetic data. The data-generation process is as follows. Let $\iota_d \in \mathbb{R}^d$ be a column vector of ones, and $I_d \in \mathbb{R}^{d \times d}$ the identity matrix. We use $\mathcal{N}(\boldmu, \boldSigma)$ to refer to a (multivariate) Gaussian distribution with mean $\boldmu \in \mathbb{R}^d$ and covariance matrix $\boldSigma \in  \mathbb{R}^{d\times d}$. The features per user for cluster $k$ are generated as:
\begin{align}
    \boldxsub{i, k} \sim \mathcal{N}(\mu_k\iota_d, I_d), \text{ where } \mu_k \sim U(-1, 1).
\end{align}
This ensures the features differ on average per user. The bias metric per user is generated as:
\begin{align}
    m_i \sim \mathcal{N}(\eta_k, 1). 
\end{align}
The $\eta_k$ is defined in line with one of two scenarios:\\
\textbf{Scenario 1: Constant bias}: the distribution of the bias metric per user is directly generated and remains consistent across clusters, i.e., $ \eta_k = 0 $ for all $ k \in \mathcal{K}$. \\
\textbf{Scenario 2: Linear increase in bias}: The expectation of the bias metric per user is directly generated as increasing from -1 to 1 (i.e., $\eta_k = -1 + 2 \cdot \frac{k-1}{K-1}$, for all $ k \in \mathcal{K}$).\\
We fix the number of clusters, $K$ and datapoints $N$, and keep the number of samples per cluster $N_k$ equal per cluster.

We fit the HBAC algorithm to this synthetic data, and afterwards test if the difference between $\barM(U\backslash U_k)$ and $\barM(U_k)$ is statistically significant at a pre-specified significance level. We highlight three recommendations for this test to have the desired properties for detecting true bias, illustrated via results on the synthetic dataset. We illustrate these recommendations when the synthetic data have $K=5$ clusters, $n=1,000$ observations, and $d=2$ features. %

In the simulation study, we identify three separate issues with the original implementation by \citet{MISZTALRADECKA2021102519} that we argued for in Section \ref{sec:tool}. Next, we outline these issues and propose a solution. 

\begin{figure}[t]
    \centering
    \includegraphics[scale=0.25]{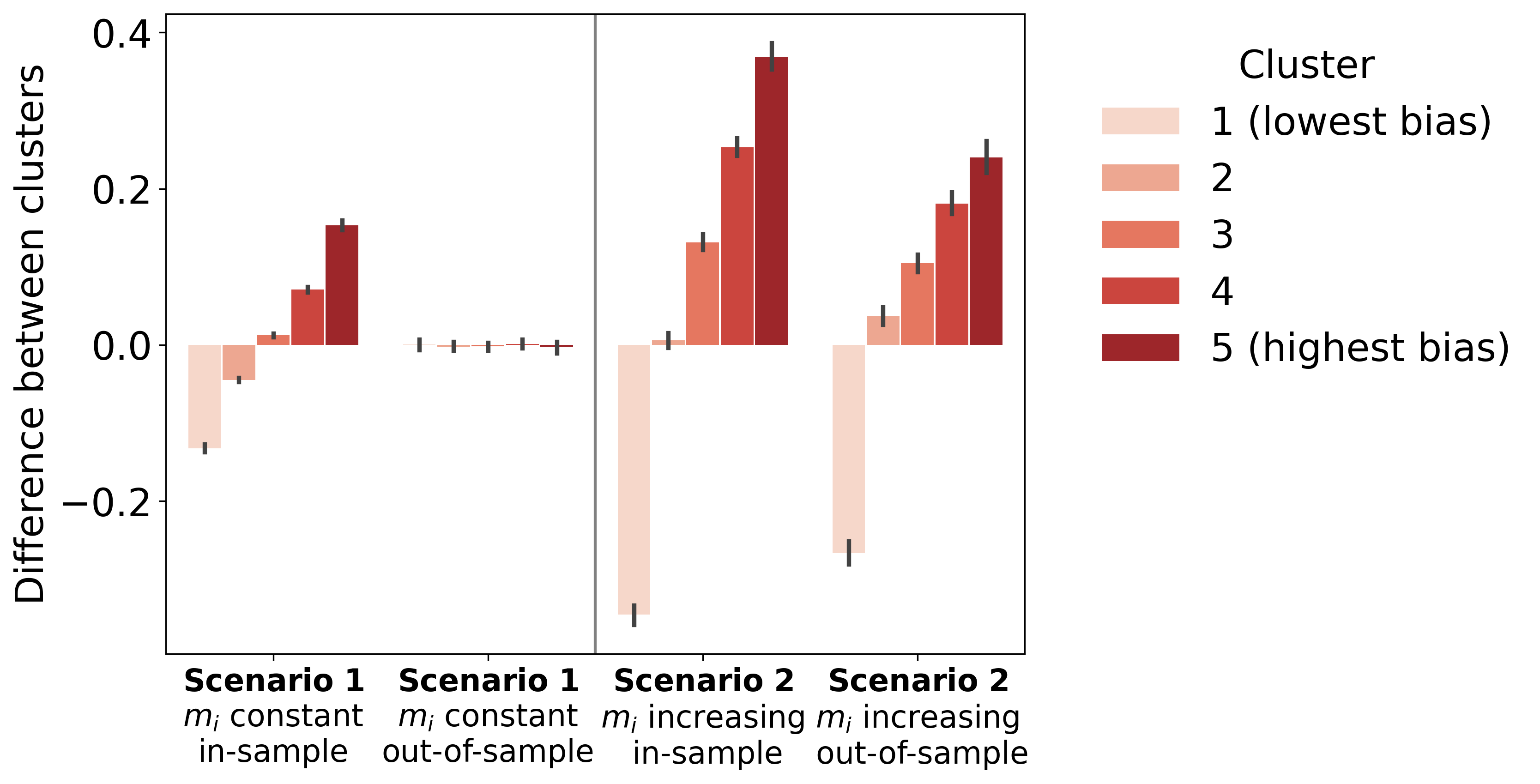}
    \caption{Average of the difference between $\barM(U\backslash U_k)$ and $\barM(U_k)$ for 1,000 simulations for the synthetic dataset with $K=5, n=1,000$, and $d=2$. The error bar reflects the 95\% confidence interval. In-sample refers to the data used to fit the HBAC algorithm, where out-of-sample refers to the data reserved for statistical testing via sample splitting. In Scenario 1 ($m_i$ constant), the expectation of $m_i$ is the same per cluster and we expect the HBAC algorithm to not find bias in the identified clusters. This is only the case when measuring the difference out-of-sample. In Scenario 2 ($m_i$ increasing), the expectation of $m_i$ linearly increases per cluster and we expect the HBAC algorithm  to find bias both in-sample and out-of-sample.}
    \label{fig:illustrate_oos}
\end{figure}

\textbf{The necessity of sample-splitting:} 
Without using sample-splitting, the difference in the bias metric is measured \textit{in-sample}, e.g. for data used to fit the HBAC algorithm. Here, there is a risk of ``overfitting.'' Figure \ref{fig:illustrate_oos} illustrates this. When the bias metric $m$ is constant for each cluster, and measured on data used to fit the HBAC algorithm, there are still substantial differences between the clusters when measuring the difference between $\barM(U\backslash U_k)$ and  $\barM(U_k)$ in-sample. However, these differences disappear when measuring the difference between $\barM(U\backslash U_k)$ and  $\barM(U_k)$ on a sample not used to fit the HBAC algorithm, e.g., \textit{out-of-sample}. This indicates that without sample splitting, we might falsely conclude that a classifier is biased. Figure \ref{fig:illustrate_oos} shows that when there is a difference in bias between the clusters, we are still able to detect this when measuring these differences out-of-sample (albeit slightly less likely).

\begin{figure*}[t]
\centering
\begin{subfigure}{.475\textwidth}
  \centering
     \includegraphics[scale=0.25]{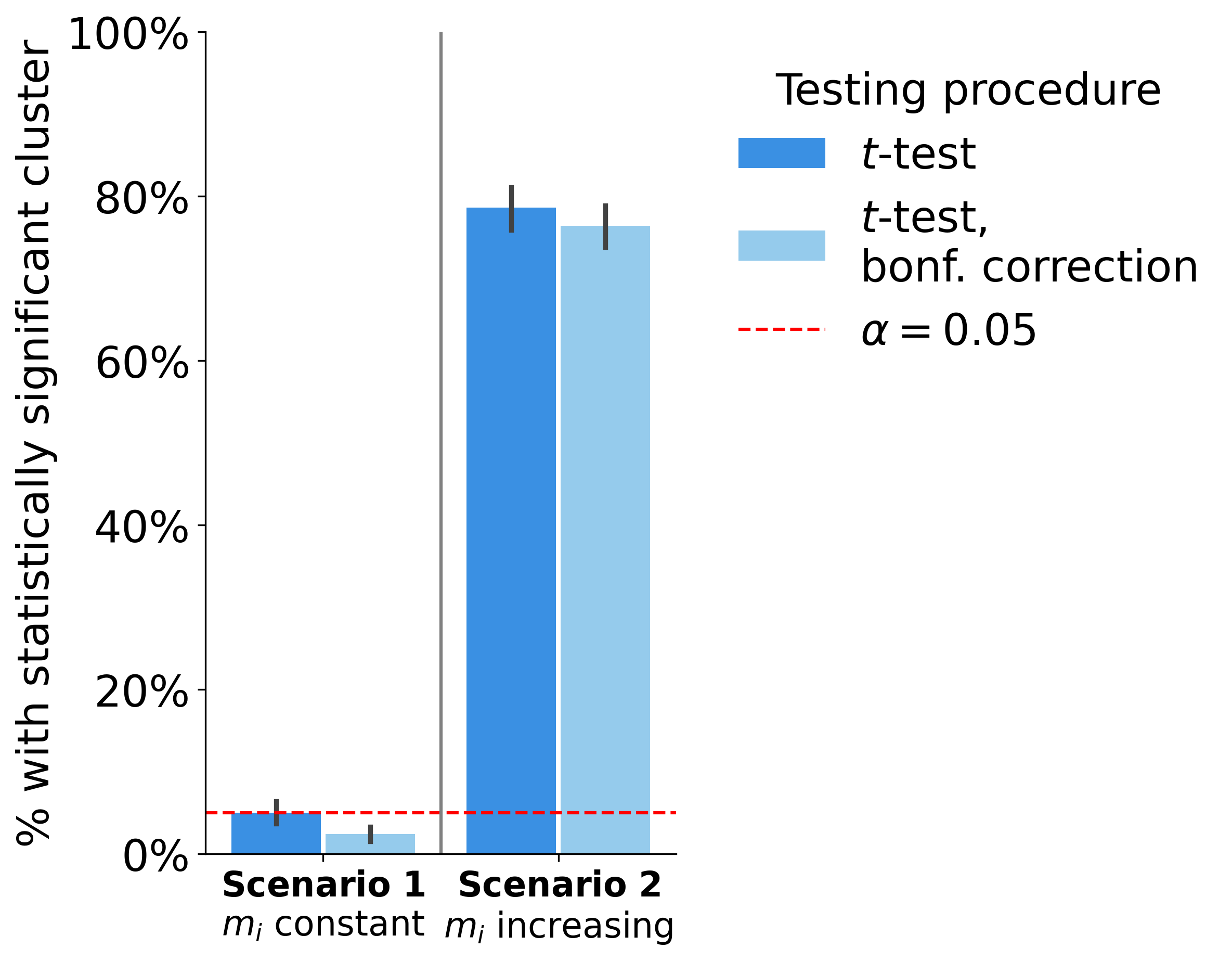} 
   \caption{The effect of using a Bonferroni correction}
  \label{fig:illustrate_bonf}
\end{subfigure}
\begin{subfigure}{.475\textwidth}
  \centering
     \includegraphics[scale=0.25]{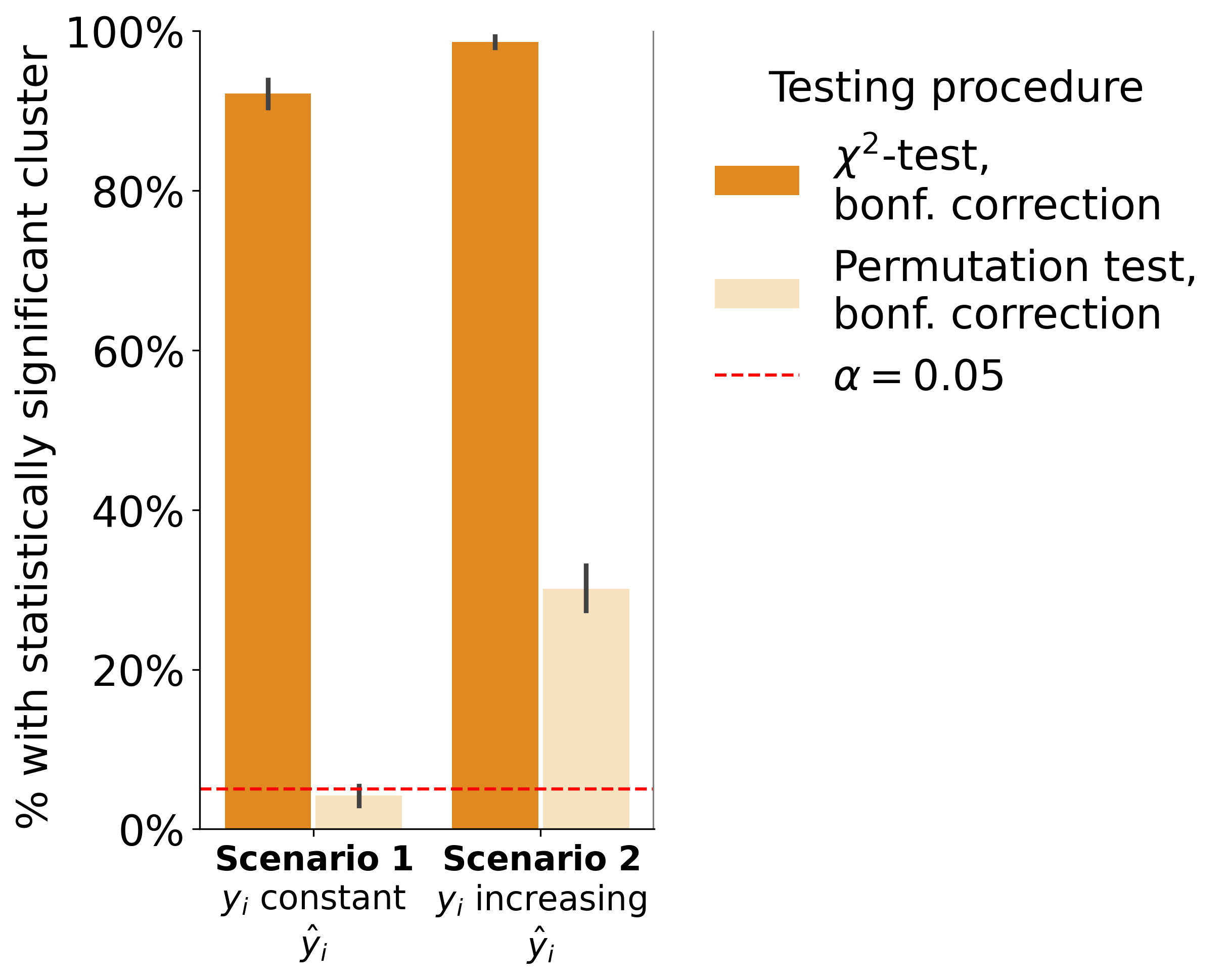} 
  \caption{The effect of using a permutation test}
  \label{fig:illustrate_perm_test}
\end{subfigure}
\caption{The percentage of simulations where for at least one of the clusters $\barM(U_k)$ is statistically significant from $\barM(U\backslash U_k)$ at the 5\% level for 1,000 simulations for the synthetic dataset with $K=5, n=1,000$ and $d=2$. The error bar reflects the 95\% confidence interval. The red dotted line reflects the significance level used when conducting the statistical test for the differences. We measure the difference between $\barM(U\backslash U_k)$ and $M(U)$ out-of-sample. In Scenario 1 the expectation of (a) $m_i$ or (b) $y_i$ is the same per cluster. In Scenario 2 the expectation of (a) $m_i$ or (b) $y_i$ linearly increases per cluster. In subfigure (b), we use $\hat{y}_i$ rather than directly simulating $m_i$, and we use $\nperm = 1,000$ permutations.}
\label{fig:multiple_perm}
\end{figure*}

\textbf{Testing across multiple clusters requires multiple testing correction:} Figure \ref{fig:illustrate_bonf} shows that without Bonferroni correction the false positive rate is much higher than the expected rate from the significance level of the test (i.e., 5\% with $\alpha = 0.05$), when there is no difference in bias between clusters. This shows that without Bonferroni correction, we might conclude that our classifier is biased, when it is not. With Bonferroni correction, the false positive rate does not exceed the expected rate (i.e., 5\%). However, this comes at a cost: Figure \ref{fig:illustrate_bonf} shows that when the bias increases per cluster, the Bonferroni correction makes it less likely that we detect differences between clusters (i.e., the true positive rate decreases).

\textbf{Different bias metrics require different null-hypotheses:}
So far, our null hypothesis has been that the difference between $\barM(U\backslash U_k)$ and $M(U)$ is zero. The validity of this null hypothesis depends on the bias metric used. For this part of the simulation study, we do not directly simulate $m_i$, but instead simulate the true attribute $y$ as follows: $ \ysub{i} = \mathrm{Bern}(p_k)$. In the first scenario, we keep  $p_k = 0.5 \ \forall k \in \mathcal{K}$, that is: the probability of $y=1$ is constant for all clusters. In the second scenario, we linearly increase the probability that $y = 1$ from 0.1 to 0.9, that is: $p_k = 0.1 + 0.8 \cdot \frac{k-1}{K-1}$. We then fit a logistic regression to predict $\ysub{i}$. Here, we only consider one bias metric: the predicted value $\hat{y}_i$ (usually used to investigate demographic parity). We consider using \emph{accuracy} as a bias metric for this experiment in Appendix \ref{sec:addendum_sim}. When using $\hat{y}_i$ as the bias metric, we observe that even when $\ysub{i}$ has the same distribution per cluster, there are differences in $\hat{y}_i$ per cluster. This is because even though $\ysub{i}$ is independent from $\boldxsub{i}$, $\hat{y}_i$ is not independent from $\boldxsub{i}$. Thus, if we use the null hypothesis that $\barM(U\backslash U_k)$ and  $\barM(U_k)$ is zero, we are likely to conclude that our classifier is biased. In Figure \ref{fig:illustrate_perm_test}, we illustrate this: if $\ysub{i}$ is constant per cluster, we incorrectly conclude there are statistically significant differences in $\hat{y}_i$. An alternative is to use the permutation test as suggested by \citet{Ojala_2010}. To do so, we permute the order of $\ysub{i}$ in our data to make it independent of $\boldxsub{i}$. By repeating this $\nperm$ times, we obtain samples from our null distribution, the distribution of the bias metric when $\boldxsub{i}$ is independent of $\ysub{i}$. We can then perform a statistical test by examining the probability of observing a value of the bias metric under the null distribution. Figure \ref{fig:illustrate_perm_test} shows the result of using such a permutation test. The $t$-test frequently suggests that there is a statistically significant difference in $\hat{y}_i$ when $\ysub{i}$ is constant. This is not true for the permutation test.

Finally, we emphasize that the appropriate null hypothesis depends on the preference of the user. It could be that a user is not interested in differences in $\hat{y}_i$ per cluster that are irreducible, e.g., that appear even if $\boldxsub{i}$ is independent of $\ysub{i}$. In this case, a permutation test is more appropriate than a $t$-test.

\begin{figure}[th]
  \centering
     \includegraphics[scale=0.25]{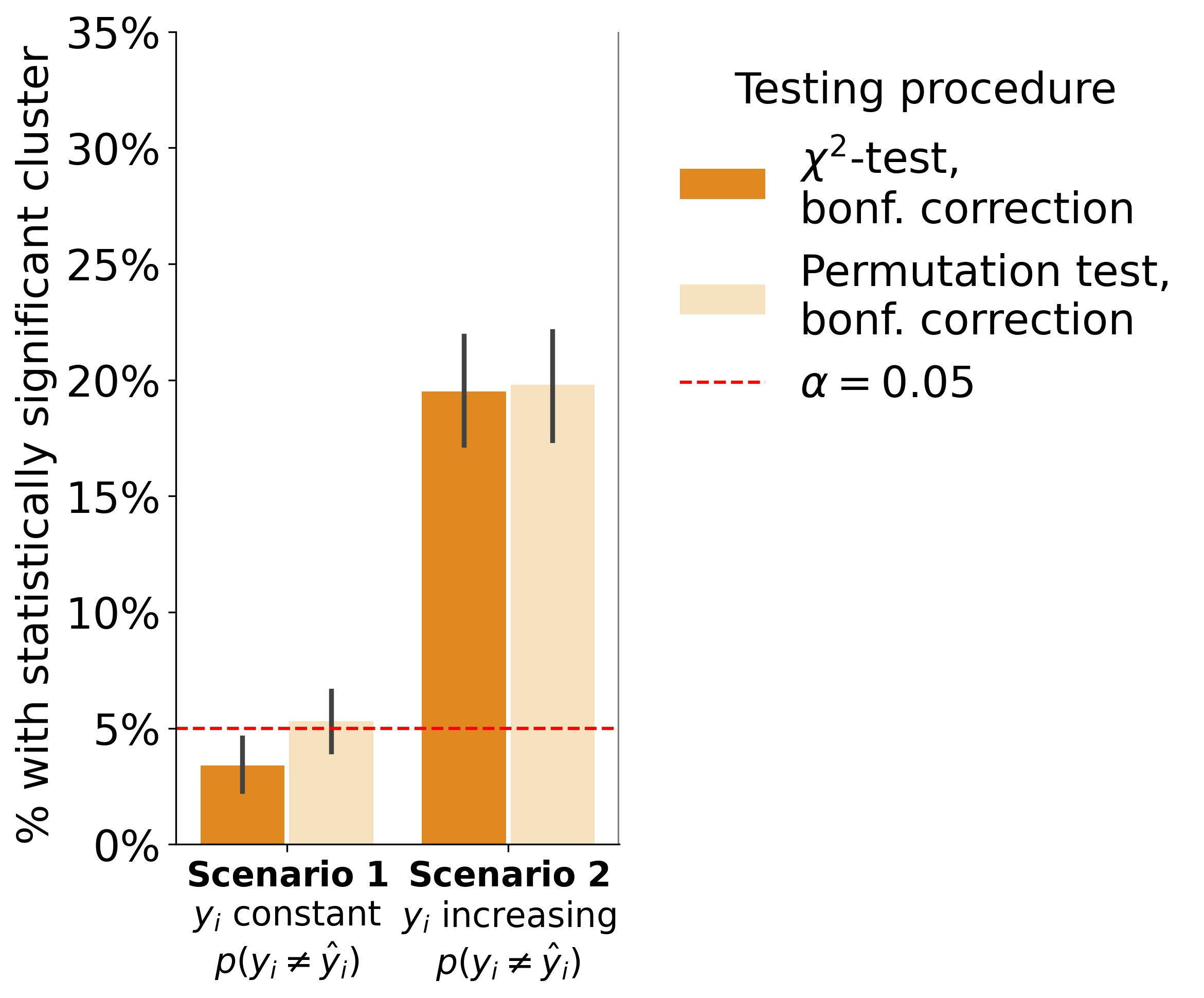} 
\caption{The percentage of simulations where for at least one of the clusters $\barM(U_k)$ is statistically significant from $\barM(U\backslash U_k)$ at the 5\% level for 1,000 simulations for the synthetic dataset with $K=5, n=1,000$ and $d=2$. The error bar reflects the 95\% confidence interval. The red dotted line reflects the significance level used when conducting the statistical test for the differences. We measure the difference between $\barM(U\backslash U_k)$ and $M(U)$ out-of-sample. In Scenario 1, the expectation of $y_i$ is the same per cluster. In Scenario 2, the expectation of  $y_i$ linearly increases per cluster. As a bias metric, we use the accuracy $p(\ysub{i} \neq \hat{y}_i)$ rather than directly simulating $m_i$, and we use $\nperm = 1,000$ permutations. }
  \label{fig:illustrate_perm_test_err}
\end{figure}

In Figure \ref{fig:illustrate_perm_test_err}, we report the results of using the permutation test when accuracy is used as a bias metric. Unlike the results for using predicted value $\hat{y}_i$ as a bias metric (see Figure \ref{fig:illustrate_perm_test}), we are not very likely to conclude there is a statistically significant difference in the bias metric. Figure \ref{fig:illustrate_perm_test_err} also suggests that using the permutation test in this scenario results in the correct probability of falsely rejecting the null hypothesis, given the significance level of 5\%.

\textbf{Limitations relating to the simulation study conducted:}
Although the simulation study is designed to be general, it relies on a relatively simple data-generation process. Future work could explore more complex scenarios, such as non-linear relationships between clusters and the bias metric, or situations where the number of data points is relatively small compared to the number of features.

\end{document}